\documentclass[letterpaper, journal, 10pt]{IEEEtran}
\usepackage{etex}

\def\papertitle{A Standalone FPGA-based Miner for Lyra2REv2 Cryptocurrencies}

\usepackage{ifpdf}
\ifpdf
\usepackage[draft,pdfborder={0 0 0}]{hyperref}
\fi

\usepackage{cite}
\usepackage{graphicx}

\usepackage[cmex10]{amsmath}
\usepackage{amssymb}
\usepackage{nicefrac}
\usepackage{bm}

\usepackage{array}
\usepackage[caption=false,font=footnotesize]{subfig}

\usepackage{tabularx}
\newcolumntype{Y}{>{\centering\arraybackslash}X}

\usepackage{multirow}
\usepackage{multicol}
\usepackage{dcolumn}
\usepackage{booktabs}

\usepackage{fixltx2e}
\usepackage{stfloats}
\usepackage{algorithm}
\usepackage{algorithmicx}
\usepackage{algpseudocode}
\algrenewcommand\algorithmicindent{1.5em}%
\usepackage{float}

\usepackage{svg}

\usepackage{color}
\usepackage{tikz,pgf}
\usepackage{pgfplots}
\usetikzlibrary{shapes,positioning,arrows,decorations.markings,fit,calc,patterns,spy,shadows}
\usetikzlibrary{plotmarks,shapes,decorations.text,decorations.pathmorphing}
\tikzstyle{branch}=[fill,shape=circle,minimum size=3pt,inner sep=0pt]

\usepackage{glossaries}

\usepackage[final]{changes}
\usepackage{hyperref}

\newcommand{\fixme}[2]{\ifx&#2&{\color{red}#1}\else{\color{red}FIXME\{}#1{\color{red}\}}\footnote{{\color{red}#2}}\PackageWarning{Fixme}{#1: #2}\fi}
\newcommand{\concat}{\:||\:}

\usepackage{xparse}
\makeatletter
\NewDocumentCommand{\LeftComment}{s m}{%
\Statex \IfBooleanF{#1}{\hspace*{\ALG@thistlm}}\(\triangleright\) #2}
\makeatother

\makeatletter
\NewDocumentCommand{\BlankComment}{s m}{%
\Statex \IfBooleanF{#1}{\hspace*{\ALG@thistlm}} #2}
\makeatother

\makeatletter

\pgfdeclareshape{reg}{
  \savedanchor\northeast{%
    \pgfmathsetlength\pgf@x{\pgfshapeminwidth}%
    \pgfmathsetlength\pgf@y{\pgfshapeminheight}%
    \pgf@x=0.325\pgf@x
    \pgf@y=0.325\pgf@y
  }
  \savedanchor\southwest{%
    \pgfmathsetlength\pgf@x{\pgfshapeminwidth}%
    \pgfmathsetlength\pgf@y{\pgfshapeminheight}%
    \pgf@x=-0.325\pgf@x
    \pgf@y=-0.325\pgf@y
  }
  \inheritanchorborder[from=rectangle]

  \anchor{center}{\pgfpointorigin}
  \anchor{north}{\northeast \pgf@x=0pt}
  \anchor{east}{\northeast \pgf@y=0pt}
  \anchor{south}{\southwest \pgf@x=0pt}
  \anchor{west}{\southwest \pgf@y=0pt}
  \anchor{north east}{\northeast}
  \anchor{north west}{\northeast \pgf@x=-\pgf@x}
  \anchor{south west}{\southwest}
  \anchor{south east}{\southwest \pgf@x=-\pgf@x}
  \anchor{text}{
    \pgfpointorigin
    \advance\pgf@x by -.5\wd\pgfnodeparttextbox%
    \advance\pgf@y by -.5\ht\pgfnodeparttextbox%
    \advance\pgf@y by +.5\dp\pgfnodeparttextbox%
  }

  \anchor{D}{
    \pgf@process{\northeast}%
    \pgf@x=-1\pgf@x%
    \pgf@y=.5\pgf@y%
  }
  \anchor{CLK}{
    \pgf@process{\northeast}%
    \pgf@x=-1\pgf@x%
    \pgf@y=-.66666\pgf@y%
  }
  \anchor{Q}{
    \pgf@process{\northeast}%
    \pgf@y=.5\pgf@y%
  }
  \backgroundpath{
    \pgfpathrectanglecorners{\southwest}{\northeast}
    \pgf@anchor@reg@CLK
    \pgf@xa=\pgf@x \pgf@ya=\pgf@y
    \pgf@xb=\pgf@x \pgf@yb=\pgf@y
    \pgf@xc=\pgf@x \pgf@yc=\pgf@y
    \pgfmathsetlength\pgf@x{0.5ex} 
    \advance\pgf@ya by \pgf@x
    \advance\pgf@xb by \pgf@x
    \advance\pgf@yc by -\pgf@x
    \pgfpathmoveto{\pgfpoint{\pgf@xa}{\pgf@ya}}
    \pgfpathlineto{\pgfpoint{\pgf@xb}{\pgf@yb}}
    \pgfpathlineto{\pgfpoint{\pgf@xc}{\pgf@yc}}
    \pgfclosepath

  }
}

\pgfdeclareshape{reg5}{
  \savedanchor\northeast{%
    \pgfmathsetlength\pgf@x{\pgfshapeminwidth}%
    \pgfmathsetlength\pgf@y{\pgfshapeminheight}%
    \pgf@x=0.325\pgf@x
    \pgf@y=0.325\pgf@y
  }
  \savedanchor\southwest{%
    \pgfmathsetlength\pgf@x{\pgfshapeminwidth}%
    \pgfmathsetlength\pgf@y{\pgfshapeminheight}%
    \pgf@x=-0.325\pgf@x
    \pgf@y=-0.325\pgf@y
  }
  \inheritanchorborder[from=rectangle]

  \anchor{center}{\pgfpointorigin}
  \anchor{north}{\northeast \pgf@x=0pt}
  \anchor{east}{\northeast \pgf@y=0pt}
  \anchor{south}{\southwest \pgf@x=0pt}
  \anchor{west}{\southwest \pgf@y=0pt}
  \anchor{north east}{\northeast}
  \anchor{north west}{\northeast \pgf@x=-\pgf@x}
  \anchor{south west}{\southwest}
  \anchor{south east}{\southwest \pgf@x=-\pgf@x}
  \anchor{text}{
    \pgfpointorigin
    \advance\pgf@x by -.5\wd\pgfnodeparttextbox%
    \advance\pgf@y by -.5\ht\pgfnodeparttextbox%
    \advance\pgf@y by +.5\dp\pgfnodeparttextbox%
  }

  \anchor{D}{
    \pgf@process{\northeast}%
    \pgf@x=-1\pgf@x%
    \pgf@y=.5\pgf@y%
  }
  \anchor{CLK}{
    \pgf@process{\northeast}%
    \pgf@x=-1\pgf@x%
    \pgf@y=-.66666\pgf@y%
  }
  \anchor{Q0}{
    \pgf@process{\northeast}%
    \pgf@y=.8\pgf@y%
  }
  \anchor{Q1}{
    \pgf@process{\northeast}%
    \pgf@y=.6\pgf@y%
  }
  \anchor{Q2}{
    \pgf@process{\northeast}%
    \pgf@y=.4\pgf@y%
  }
  \anchor{Q3}{
    \pgf@process{\northeast}%
    \pgf@y=.2\pgf@y%
  }
  \anchor{Q4}{
    \pgf@process{\northeast}%
    \pgf@y=0\pgf@y%
  }
  
  \backgroundpath{
    \pgfpathrectanglecorners{\southwest}{\northeast}
    \pgf@anchor@reg@CLK
    \pgf@xa=\pgf@x \pgf@ya=\pgf@y
    \pgf@xb=\pgf@x \pgf@yb=\pgf@y
    \pgf@xc=\pgf@x \pgf@yc=\pgf@y
    \pgfmathsetlength\pgf@x{0.5ex} 
    \advance\pgf@ya by \pgf@x
    \advance\pgf@xb by \pgf@x
    \advance\pgf@yc by -\pgf@x
    \pgfpathmoveto{\pgfpoint{\pgf@xa}{\pgf@ya}}
    \pgfpathlineto{\pgfpoint{\pgf@xb}{\pgf@yb}}
    \pgfpathlineto{\pgfpoint{\pgf@xc}{\pgf@yc}}
    \pgfclosepath

  }
}

\tikzset{add font/.code={\expandafter\def\expandafter\tikz@textfont\expandafter{\tikz@textfont#1}}} 

\tikzset{flip flop/port labels/.style={font=\sffamily\scriptsize}}
\tikzset{every reg node/.style={draw,fill=blue!10,minimum width=0.75cm,minimum height=2.2cm,thick,inner sep=1mm,outer sep=0pt,cap=round,add font=\sffamily\scriptsize}}
\tikzset{every reg5 node/.style={draw,fill=blue!10,minimum width=0.75cm,minimum height=5.1cm,thick,inner sep=1mm,outer sep=0pt,cap=round,add font=\sffamily\scriptsize}}

\makeatother

\makeatletter

\tikzset{every mux5invsel node/.style={draw,minimum width=0.5cm,minimum height=2.2cm,inner sep=1mm,outer sep=0pt}}

\pgfdeclareshape{mux5invsel}{

  \savedanchor\northeast{%
    \pgfmathsetlength\pgf@x{\pgfshapeminwidth}%
    \pgfmathsetlength\pgf@y{\pgfshapeminheight}%
    \pgf@x=0.5\pgf@x
    \pgf@y=0.5\pgf@y
  }
  \savedanchor\southwest{%
    \pgfmathsetlength\pgf@x{\pgfshapeminwidth}%
    \pgfmathsetlength\pgf@y{\pgfshapeminheight}%
    \pgf@x=-0.5\pgf@x
    \pgf@y=-0.5\pgf@y
  }
  \inheritanchorborder[from=rectangle]
  
  \anchor{center}{\pgfpointorigin}
  \anchor{north}{\northeast \pgf@x=0pt}
  \anchor{east}{\northeast \pgf@y=0pt}
  \anchor{south}{\southwest \pgf@x=0pt}
  \anchor{west}{\southwest \pgf@y=0pt}
  \anchor{north east}{
    \pgf@process{\northeast}%
    \pgf@y=.25\pgf@y%
  }
  \anchor{north west}{\northeast \pgf@x=-\pgf@x}
  \anchor{south west}{\southwest}
  \anchor{south east}{\southwest \pgf@x=-\pgf@x}
  \anchor{text}{
    \pgfpointorigin
    \advance\pgf@x by -.5\wd\pgfnodeparttextbox%
    \advance\pgf@y by -.5\ht\pgfnodeparttextbox%
    \advance\pgf@y by +.5\dp\pgfnodeparttextbox%
  }

  \anchor{in0}{
    \pgf@process{\northeast}%
    \pgf@x=-1\pgf@x%
    \pgf@y=0.6\pgf@y%
  }
  \anchor{in1}{
    \pgf@process{\northeast}%
    \pgf@x=-1\pgf@x%
    \pgf@y=0.3\pgf@y%
  }
  \anchor{in2}{
    \pgf@process{\northeast}%
    \pgf@x=-1\pgf@x%
    \pgf@y=0\pgf@y%
  }
  \anchor{in3}{
    \pgf@process{\northeast}%
    \pgf@x=-1\pgf@x%
    \pgf@y=-0.3\pgf@y%
  }
  \anchor{in4}{
    \pgf@process{\northeast}%
    \pgf@x=-1\pgf@x%
    \pgf@y=-0.6\pgf@y%
  }
  \anchor{sel}{
    \pgf@process{\northeast}%
    \pgf@x=0pt%
    \pgf@y=0.77\pgf@y%
  }

  \anchor{out}{
    \pgf@process{\northeast}%
    \pgf@y=0pt%
  }

  \backgroundpath{
    \southwest \pgf@xa=\pgf@x \pgf@ya=\pgf@y
    \northeast \pgf@xb=\pgf@x \pgf@yb=\pgf@y
    \let\pgf@xnw=\pgf@xa
    \let\pgf@ynw=\pgf@yb
    \let\pgf@xsw=\pgf@xa
    \let\pgf@ysw=\pgf@ya
    \let\pgf@xse=\pgf@xb
    \pgf@yc=-\pgf@yb
    \advance\pgf@yc by \pgfshapeminwidth
    \let\pgf@yse=\pgf@yc
    \let\pgf@xne=\pgf@xb
    \pgf@xc=\pgf@yb
    \advance\pgf@xc by -\pgfshapeminwidth
    \let\pgf@yne=\pgf@xc
    \pgfpathmoveto{\pgfpoint{\pgf@xnw}{\pgf@ynw}}
    \pgfpathlineto{\pgfpoint{\pgf@xsw}{\pgf@ysw}}
    \pgfpathlineto{\pgfpoint{\pgf@xse}{\pgf@yse}}
    \pgfpathlineto{\pgfpoint{\pgf@xne}{\pgf@yne}}
    \pgfpathclose
  }
}

\makeatother

\title{\papertitle}

\author{Jean-Fran\c{c}ois T\^{e}tu$^*$, Louis-Charles Trudeau$^*$, Michiel Van Beirendonck$^*$, \\Alexios Balatsoukas-Stimming,~\IEEEmembership{Member,~IEEE,} and Pascal Giard,~\IEEEmembership{Senior Member,~IEEE}
  \thanks{$^*$Equally contributing authors presented in alphabetical order.}%
  \thanks{J.-F. T\^{e}tu and L.-C. Trudeau were with and P. Giard is with the LaCIME, \'Ecole de technologie sup\'erieure (ETS), Montreal, QC, Canada (e-mails: jftetu@jftetu.net, louis.charles.trudeau@gmail.com, pascal.giard@etsmtl.ca).}%
	\thanks{M. Van Beirendonck is with imec-COSIC KU Leuven, Leuven, Belgium (e-mail: michiel.vanbeirendonck@esat.kuleuven.be).}%
	\thanks{A. Balatsoukas-Stimming is with the Telecommunications Circuits Laboratory, \'Ecole polytechnique f\'ed\'erale de Lausanne (EPFL), Lausanne, VD, Switzerland and with the Department of Electrical Engineering, Eindhoven University of Technology, Eindhoven, The Netherlands (e-mail: a.k.balatsoukas.stimming@tue.nl).}%
	\thanks{Parts of this work were presented at the 2019 IEEE International Symposium on Circuits and Systems in Sapporo, Japan~\cite{VanBeirendonck_ISCAS_2019}.}%
}

\begin{document}

\newacronym{kdf}{KDF}{key derivation function}
\newacronym{phs}{PHS}{password hashing scheme}
\newacronym{fsm}{FSM}{finite-state machine}
\newacronym[firstplural=D flip-flops (DFFs)]{dff}{DFF}{D flip-flop}
\newacronym{bram}{BRAM}{block RAM}
\newacronym{bmw}{BMW}{Blue Midnight Wish}
\newacronym{pow}{PoW}{proof-of-work}
\newacronym{pos}{PoS}{proof-of-stake}
\newacronym{pob}{PoB}{proof-of-burn}
\newacronym{cc}{CC}{clock cycle}
\newacronym{mux}{MUX}{multiplexer}
\newacronym[firstplural=Multi-Processor System on Chips (MPSoCs)]{mpsoc}{MPSoC}{Multi-Processor System on Chip}
\newacronym{pl}{PL}{programmable logic}
\newacronym{ps}{PS}{processing system}
\newacronym{bsp}{BSP}{board support package}
\newacronym{ubi}{UBI}{unique block iteration}
\newacronym{gmu}{GMU}{George Mason University}
\newacronym{lut}{LUT}{look-up table}
\newacronym{clb}{CLB}{configurable logic block}

\maketitle

\begin{abstract}
Lyra2REv2 is a hashing algorithm that consists of a chain of individual hashing algorithms, and it is used as a \glsentrylong{pow} function in several cryptocurrencies. The most crucial and exotic hashing algorithm in the Lyra2REv2 chain is a specific instance of the general Lyra2 algorithm. This work presents the first hardware implementation of the specific instance of Lyra2 that is used in Lyra2REv2. Several properties of the aforementioned algorithm are exploited in order to optimize the design. In addition, an FPGA-based hardware implementation of a standalone miner for Lyra2REv2 on a Xilinx \Glsentrylong{mpsoc} is presented. The proposed Lyra2REv2 miner is shown to be significantly more energy efficient than both a GPU and a commercially available FPGA-based miner. Finally, we also explain how the simplified Lyra2 and Lyra2REv2 architectures can be modified with minimal effort to also support the recent Lyra2REv3 chained hashing algorithm.
\end{abstract}

\begin{IEEEkeywords} Lyra2, Lyra2REv2, Lyra2REv3, hardware miner, FPGA miner, MPSoC miner, cryptocurrency\end{IEEEkeywords}

\section{Introduction}
\label{sec:intro}
\IEEEPARstart{R}{ecently}, there has been a surge in the popularity of cryptocurrencies, which are digital currencies that enable transactions through a decentralized consensus mechanism. Most cryptocurrencies are based on a \emph{blockchain}, which is an ever-growing list of transactions that are grouped in blocks. Individual blocks in the chain are linked together using a cryptographic hash of the previous block, which ensures resistance against modifications, and every transaction is digitally signed, typically by using public-key cryptography. Various mechanisms are used in order to deter denial-of-service attacks and, in particular, \emph{double-spending} attacks where the same digital coin is used in multiple concurrent transactions. Many popular cryptocurrencies, incuding Bitcoin~\cite{bitcoin}, use a \gls{pow} mechanism, which was first proposed in~\cite{Dwork1993} to combat the problem of junk mail. The \gls{pos} and \gls{pob} mechanisms are other notable proposals.

The \gls{pow} system requires that new blocks provide proof that a function that requires a significant amount of a limited resource was used to construct them before they get accepted into the chain. For example, the employed function can be limited by the available processing power, the available memory, or the network bandwidth and latency. Cryptocurrencies typically use functions that are limited by the available processing power, the most common approach being that random numbers are appended to a block until its cryptographic hash meets a certain condition (e.g., some of its most-significant bits are equal to $0$). The chain with the most cumulative \gls{pow} is accepted as the correct one, so that an attacker must control more than half of the active processing power on the network to perform a double-spend attack. This is unlikely to happen in practice if the processing power is large enough and is owned by non-colluding entities. Processing nodes that help to compute the hashes of new blocks are called \emph{miners}, and are rewarded with a fraction of the cryptocurrency when a new block is accepted into the blockchain.

The first cryptocurrency, i.e., Bitcoin~\cite{bitcoin}, was initially mined using desktop CPUs. Then, GPUs were used to significantly increase the hashing speed. Eventually, GPU mining was outpaced by FPGA miners, which were in turn surpassed by ASIC miners. Nowadays, the majority of the computing power on the Bitcoin network is found in large ASIC farms, each operated by a single entity, which makes the decentralized nature of Bitcoin debatable. To solve this issue, new~\gls{pow} algorithms have been proposed that aim to be ASIC-resistant. ASIC resistance is achieved by using hashing algorithms that are highly serial, memory-intensive, and parameterizable so that a manufactured ASIC can easily be made obsolete by changing some of the parameters. Since the cost of manufacturing new ASICs whenever some parameters change is prohibitive, GPU mining of ASIC-resistant cryptocurrencies is generally much more low-risk and cost-effective. A prime example of an ASIC-resistant hashing algorithm is Lyra2REv2 (and its recently introduced Lyra2REv3 modification), which is used by MonaCoin~\cite{monacoin}, Verge~\cite{verge}, Vertcoin~\cite{vertcoin}, and some smaller cryptocurrencies. The chained structures of Lyra2REv2 and Lyra2REv3 are shown in Fig.~\ref{fig:Lyra2REv2} and Fig.~\ref{fig:Lyra2REv3}, respectively. The BLAKE~\cite{Aumasson2008}, Keccak~\cite{Bertoni2011}, Skein~\cite{Ferguson2010}, \gls{bmw}~\cite{Gligoroski2009}, and CubeHash~\cite{Bernstein2009} hashing algorithms are well-known and have been studied heavily, both from theoretical and hardware-implementation perspectives (e.g., \cite{Tillich2009,Baldwin2010,Gaj2010,gmu}), as they were all candidates in the SHA-3 competition.
On the other hand, to the best of our knowledge, apart from our own previous work~\cite{VanBeirendonck_ISCAS_2019}, no hardware implementation of the simplified Lyra2 and Lyra2MOD versions of Lyra2~\cite{lyra2,lyra2TC} as used in the Lyra2REv2 and Lyra2REv3 algorithms, respectively, have been reported in the literature.

One potential issue with ASIC-resistant cryptocurrencies is that GPUs are generally much less energy efficient than ASICs, meaning that a massive adoption of ASIC-resistant cryptocurrencies would significantly increase the (already very high) energy consumption of cryptocurrency mining. FPGA-based miners, on the other hand, are flexible, energy efficient, and readily available to the general public at reasonable prices. Thus, provided that public and user-friendly FPGA-based miners become available, we believe that FPGAs are in fact an attractive platform for ASIC-resistant cryptocurrencies that should not be shunned by the community.

\subsubsection*{Contributions}
This work presents the first FPGA implementation of the simplified Lyra2 hashing algorithm as used in Lyra2REv2. Moreover, contrary to our previous work \cite{VanBeirendonck_ISCAS_2019} which only contained an implementation of the Lyra2 core, in this work we describe an FPGA-based hardware implementation of a \emph{fully functional standalone Lyra2REv2 miner} on a Xilinx \gls{mpsoc}. While we do not provide explicit implementation results for Lyra2MOD or for a Lyra2REv3 chain, which is currently only used by the (somewhat less popular) Vertcoin cryptocurrency, we explain in detail how the presented architecture can be modified correspondingly. We present post-layout results for a Xilinx \gls{mpsoc} for the complete standalone Lyra2REv2 miner as well as for the individual hashing cores. These results show that the proposed Lyra2REv2 hardware architecture can achieve a hashing throughput of 31.25\,MHash/s with an energy efficiency that is up to 4.3 times better than existing solutions at 0.80\,$\mu$J/Hash, while requiring approximately 85\% of the \gls{pl} resources of the~\gls{mpsoc}.

\subsubsection*{Outline}
The remainder of this paper is organized as follows. Section~\ref{sec:bg} provides the necessary background for the \gls{pow} concept and for the Lyra2 algorithm. Section~\ref{sec:SimpleLyra2} gives an in-depth explanation of the simplifications that Lyra2REv2 and Lyra2REv3 make to the generic Lyra2 algorithm. The hardware implementations of the simplified Lyra2 and Lyra2MOD algorithms are described at length in Section~\ref{sec:impl:SimpleLyra2}. Section~\ref{sec:impl:Lyra2REv2} describes an \gls{mpsoc}-based hardware architecture that implements the Lyra2REv2 miner, which can be easily modified to also implement a Lyra2REv3 miner. Implementation details and results for the standalone Lyra2REv2 miner are provided in Section~\ref{sec:results}. Section~\ref{sec:results} also includes results for the individual hashing cores, notably including the proposed simplified Lyra2 core. Finally, Section~\ref{sec:conclusion} concludes this paper.

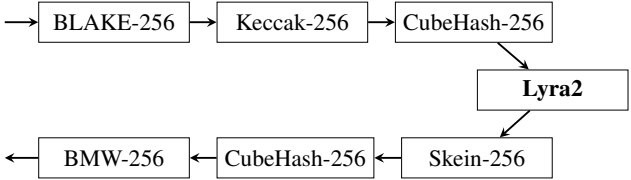
\begin{figure}[t]
  \centering
  \resizebox{0.95\columnwidth}{!}{
  \begin{tikzpicture}[node distance=0.4cm]
    \tikzstyle{hash} = [rectangle, minimum width=2.25cm, minimum height=0.6cm,text centered, draw=black]
    \tikzstyle{arrow} = [thick,->,>=stealth]

    \node (blake) [hash] {BLAKE-256};
    \node (keccak) [hash,right=of blake.east] {Keccak-256};
    \node (cube1) [hash,right=of keccak.east] {CubeHash-256};
    \node (lyra2) [hash,below=of cube1.south east] {\bf Lyra2};
    \node (skein) [hash,below=of lyra2.south west] {Skein-256};
    \node (cube2) [hash,left=of skein.west] {CubeHash-256};
    \node (bmw) [hash,left=of cube2.west] {BMW-256};

    \draw [arrow] ([xshift=-0.5cm]blake.west) -- (blake);
    \draw [arrow] (blake) -- (keccak);
    \draw [arrow] (keccak) -- (cube1);
    \draw [arrow] (cube1) -- (lyra2);
    \draw [arrow] (lyra2) -- (skein);
    \draw [arrow] (skein) -- (cube2);
    \draw [arrow] (cube2) -- (bmw);
    \draw [arrow] (bmw) -- ([xshift=-0.5cm]bmw.west);
  \end{tikzpicture}}%
  \caption{The Lyra2REv2 chained hashing algorithm.}\vspace{-0.8em}
  \label{fig:Lyra2REv2}
\end{figure}

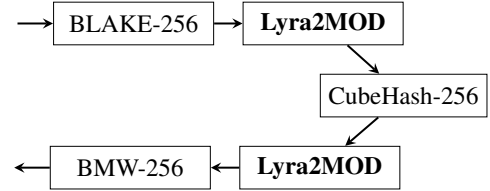
\begin{figure}[t]
  \centering
  \resizebox{0.725\columnwidth}{!}{
  \begin{tikzpicture}[node distance=0.4cm]
    \tikzstyle{hash} = [rectangle, minimum width=2.25cm, minimum height=0.6cm,text centered, draw=black]
    \tikzstyle{arrow} = [thick,->,>=stealth]

    \node (blake) [hash] {BLAKE-256};
    \node (lyra2-1) [hash,right=of blake.east] {\bf Lyra2MOD};
    \node (cube) [hash,below=of lyra2-1.south east] {CubeHash-256};
    \node (lyra2-2) [hash,below=of cube.south west] {\bf Lyra2MOD};
    \node (bmw) [hash,left=of lyra2-2.west] {BMW-256};

    \draw [arrow] ([xshift=-0.5cm]blake.west) -- (blake);
    \draw [arrow] (blake) -- (lyra2-1);
    \draw [arrow] (lyra2-1) -- (cube);
    \draw [arrow] (cube) -- (lyra2-2);
    \draw [arrow] (lyra2-2) -- (bmw);
    \draw [arrow] (bmw) -- ([xshift=-0.5cm]bmw.west);
  \end{tikzpicture}}%
  \caption{The Lyra2REv3 chained hashing algorithm.}
  \label{fig:Lyra2REv3}
\end{figure}

\section{Background}
\label{sec:bg}
This section provides the necessary background on the \gls{pow} concept, as well as some components of the Keccak and BLAKE2 hashing algorithms which are used in Lyra2.

\subsection{Proof of Work}\label{sec:bg:pow}
In order to explain the \gls{pow} concept in more detail, we use Bitcoin as an example~\cite{bitcoinreference}, but it is important to note that many other Bitcoin-derived cryptocurrencies, such as MonaCoin and Vertcoin, use the same structure. Each block in the Bitcoin blockchain has an $80$-byte (or $640$-bit) header that contains information about the block, as shown in Table~\ref{tab:bcheader}. The \texttt{version} field dictates which version of the block validation rules needs to be followed. The \texttt{previous block header hash} and \texttt{merkle root hash} contain hashes of the headers of previous blocks to ensure that no previous transaction in the blockchain can be modified without also modifying the header of the current block. The \texttt{time} field contains the Unix epoch at which each miner started performing the~\gls{pow}, which must be strictly greater than the median time of the previous $11$ blocks. The \texttt{nBits} and \texttt{nonce} fields are the most relevant to the \gls{pow}. Specifically, \texttt{nBits} defines a $256$-bit numerical value using an encoding explained in~\cite{bitcoinreference}, while \texttt{nonce} can be chosen freely by the miner. The \gls{pow} that each miner performs amounts to finding a value for \texttt{nonce} so that a (chained) hash function of the header has a numerical value that is strictly smaller than the target threshold defined by \texttt{nBits}. Since hash functions possess the property of preimage resistance, i.e., they are not invertible, this can only be achieved by testing a very large number of \texttt{nonce} values until the target threshold is satisfied.

\begin{table}[t]
	\centering
	\caption{Contents of the Bitcoin Block Header}
	\label{tab:bcheader}
	\begin{tabular}{c|c}
		\toprule\hline
		Bytes		& Name \\
		\hline\hline
		$4$			& \texttt{version} \\
		$32$		& \texttt{previous block header hash} \\
		$32$		& \texttt{merkle root hash} \\
		$4$			& \texttt{time} \\
		$4$			& \texttt{nBits} \\
		$4$			& \texttt{nonce} \\
		\hline\bottomrule
	\end{tabular}
\end{table}

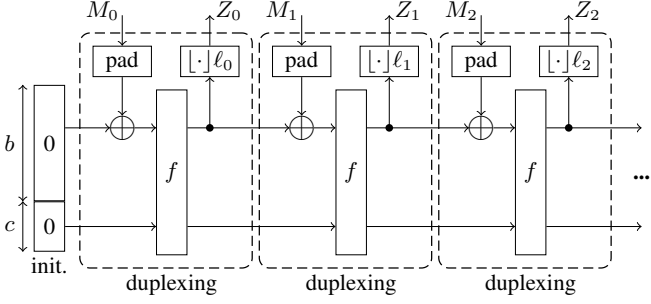
\begin{figure}
	\centering
	\resizebox{\columnwidth}{!}{
    \begin{tikzpicture}[%
      font=\small,%
      inner sep=0pt,%
      minimum width=1.0em,%
      square/.style={regular polygon,regular polygon sides=4},%
      ]

      \tikzset{black dotted/.style={draw=black, line width=0.5pt,
          dash pattern=on 4pt off 2pt,
          inner sep=0.5em, rectangle, rounded corners}};

      \def\nsep{1.5em}
      \def\fwidth{1.2em}
      \def\fheight{6.5em}
      \def\owidth{2.3em}
      \def\os{-0.17}

      \node[draw=black,minimum width=\fwidth, minimum height=0.7*\fheight] (stateb) at (0,0) {0};
      \node[draw=black,anchor=north, minimum width=\fwidth, minimum height=0.3*\fheight] (statec) at (stateb.south) {0}; 
      \node[anchor=north,yshift=-.2em] at (statec.south) {init.};

      \draw[<->] ($(stateb.north west)-(0.15,0)$) -- node[left] {$b$} ($(stateb.south west)-(0.15,0)$);
      \draw[<->] ($(statec.north west)-(0.15,0)$) -- node[left] {$c$} ($(statec.south west)-(0.15,0)$);

      \node (op) at ($(stateb.east)+(1.5*\nsep,0.4*\nsep)$) {\Large $\oplus$};
      \node[draw=black,anchor=west, minimum width=\fwidth, minimum height=\fheight] (f) at ($(op.east)+(0.5*\nsep,-0.27*\fheight)$) {$f$};
      \node[anchor=north,yshift=-0.75em] (duplex.south) at (f.south) {duplexing};
      \node[draw=black,anchor=south, minimum width=\owidth, minimum height=\fwidth] (pad) at ($(op)+(0,2.0em)$) {pad}; 
      \node[draw=black,anchor=south, minimum width=\owidth, minimum height=\fwidth] (hash) at ($(f |- op)+(\nsep,2.0em)$) {$\lfloor\cdot\rfloor \ell_0$}; 

      \draw[->] (stateb.east |- op) -- ($(op)+(\os,0)$);
      \draw[->] (pad.south) -- ($(op)+(0,-\os)$);
      \draw[->] ($(op)-(\os,0)$) -- (f.west |- op);
      \draw[->] (statec.east) -- (f.west |- statec);
      \draw[->] (op -| hash) node[branch] {} -- (hash.south);

      \draw[->] ($(pad.north)+(0,0.8*\nsep)$) -- node[anchor=south east,xshift=-0.1em,yshift=0.4em] {$M_0$} (pad.north);
      \draw[<-] ($(hash.north)+(0,0.8*\nsep)$) -- node[anchor=south west,xshift=0.2em,yshift=0.4em] {$Z_0$} (hash.north);

      \node (duplex) [black dotted, fit = (pad) (hash) (f)] {};
      
      \foreach \x in {1, 2}
      {
        \node (op) at ($(f.east |- op)+(3*\nsep,0)$) {\Large $\oplus$};
        \draw[->] (f.east |- op) -- ($(op)+(\os,0)$);
        \draw[->] (f.east |- statec) -- ($(op.east |- statec)+(0.5*\nsep,0)$);
        \node[draw=black,anchor=west, minimum width=\fwidth, minimum height=\fheight] (f) at ($(op.east)+(0.5*\nsep,-0.27*\fheight)$) {$f$}; 
        \node[anchor=north,yshift=-0.75em] (duplex.south) at (f.south) {duplexing};
        \node[draw=black,anchor=south, minimum width=\owidth, minimum height=\fwidth] (pad) at ($(op)+(0,2.0em)$) {pad}; 
        \node[draw=black,anchor=south, minimum width=\owidth, minimum height=\fwidth] (hash) at ($(f |- op)+(\nsep,2.0em)$) {$\lfloor\cdot\rfloor \ell_\x$}; 

        \draw[->] (pad.south) -- ($(op)+(0,-\os)$);
        \draw[->] ($(op)-(\os,0)$) -- (f.west |- op);
        \draw[->] (op -| hash) node[branch] {} -- (hash.south);

        \draw[->] ($(pad.north)+(0,0.8*\nsep)$) -- node[anchor=south east,xshift=-0.1em,yshift=0.4em] {$M_\x$} (pad.north);
        \draw[<-] ($(hash.north)+(0,0.8*\nsep)$) -- node[anchor=south west,xshift=0.2em,yshift=0.4em] {$Z_\x$} (hash.north);

        \node (duplex) [black dotted, fit = (pad) (hash) (f)] {};
      }

      \draw[->] (f.east |- op) -- ($(f.east |- op)+(2.5*\nsep,0)$);
      \draw[->] (f.east |- statec) -- ($(f.east |- statec)+(2.5*\nsep,0)$);
      \node (cdots) at ($(f.east |- op)!0.5!(f.east |- statec)+(2.5*\nsep,0)$) {\textbf{...}};

    \end{tikzpicture}
  }%
	\caption{The Keccak-based duplex construction as adopted in Lyra2 (reproduced based on~\cite[Fig. 2]{lyra2refguide}).}\label{fig:Lyra2KeccakDuplex}
	
\end{figure}

\subsection{The Keccak Duplex}\label{sec:bg:keccak}
Keccak is a family of hashing algorithms based on a cryptographic sponge \cite{sponge,sha3}. A cryptographic sponge is a function that takes an arbitrary-length input to produce an arbitrary-length hashed output. Lyra2 uses a specific implementation of the sponge, called the \emph{duplex} construction, which has a state that is preserved across different inputs. The duplex construction with naming conventions as adopted in Lyra2 is shown in Fig.~\ref{fig:Lyra2KeccakDuplex}. It consists of a permutation function $f$ that operates on a $w$-bit state vector, where $w=b+c$ and the parameters $b$ and $c$ are called the \emph{bitrate} and the \emph{capacity} of the sponge, respectively, as well as a padding rule \texttt{pad}. Note that the permutation $f$ is iterative and performs a pre-defined number of iterations, also called \emph{rounds}.

A call to the duplex construction proceeds as follows. An input string $M$ is first fed into the duplex. Then, it is padded to length $b$ and XOR'd into the lower $b$ bits of the state. The state is then fed through the permutation $f$. The output of $f$ is the new state of the duplex, while its lower $l$ bits are the output hash, where $l \leq b$. If the duplex construction is considered as an object $H$, then the aforementioned procedure is referred to as a method $H.\texttt{duplex}(M,l)$. The following two auxiliary methods are useful to simplify the notation: $H.\texttt{absorb}(M)$ updates the state using the input $M$ but discards the output (equivalent to $H.\texttt{duplex}(M,0)$), while $H.\texttt{squeeze}(l)$ reads $l$ output bits and then calls $H.\texttt{absorb}(\emptyset)$, where $\emptyset$ denotes an empty input string.

\subsection{The BLAKE2b Round Function}\label{sec:bg:blake2b}
BLAKE2 \cite{blake2} is a family of hash functions designed for fast software implementations. It is the successor of BLAKE as submitted to the SHA-3 competition \cite{blake}.
The Lyra2 algorithm heavily draws from the round function of BLAKE2b, the 64-bit variant of BLAKE2.
The round function consists of an arrangement of blocks that apply a so-called G-function to a 16-word state, where one G-function operates on 4 different state words.
For BLAKE2b a word has 64 bits meaning that 16 state words amount to 1024 bits. The total round transforms these 1024 bits using four G-blocks, rearranges the output, and then does a four G-block transformation again. Algorithm\,\ref{algo:blake2b:g} describes the modified BLAKE2b G-function as used in Lyra2, where $x \ggg y$ denotes a $y$-bit right rotation of $x$ and $\boxplus$ denotes a word-wise modulo-$2^m$ addition where the word width of both the operands and the result is $m$ bits.

\begin{algorithm}[t]
  \caption{The G-function of BLAKE2b as used in Lyra2}
  \label{algo:blake2b:g}
  \begin{algorithmic}[1]\small
    \State\texttt{INPUTS:} $a,b,c,d$
    \State \texttt{OUTPUTS:} $a',b',c',d'$
    \State $a'\gets a \boxplus b$
    \State $d'\gets (d \oplus a') \ggg 32$
    \State $c'\gets c \boxplus d'$
    \State $b'\gets (b \oplus c') \ggg 24$
    \State $a'\gets a' \boxplus b'$
    \State $d'\gets (d' \oplus a') \ggg 16$
    \State $c'\gets c' \boxplus d'$
    \State $b'\gets (b' \oplus c') \ggg 63$
  \end{algorithmic}
\end{algorithm}

\section{The Simplified Lyra2 Algorithms Used in Lyra2REv2 and Lyra2REv3}\label{sec:SimpleLyra2}

\begin{algorithm}[t]
  \caption{Lyra2 algorithm as specified in Lyra2REv2.}
  \label{algo:lyra2}
  \begin{algorithmic}[1]\fontsize{8.75pt}{8.75pt}\selectfont
   	\State \texttt{PARAMS:} $H, \rho, \omega, T, R, C, k, b\text{ as }H.b$
    \State\texttt{INPUT:} $pwd$
    \State \texttt{OUTPUT:} $K$

    \vspace*{0.3em}
    \LeftComment{\textbf{Bootstrapping Phase}}

    \State $params \gets \texttt{len}(K) \concat \texttt{len}(pwd) \concat \texttt{len}(pwd) \concat T \concat R \concat C$
    \State $H.\texttt{absorb}(\texttt{pad}(pwd \concat pwd \concat params))$  \label{lst:line:boostrap}

    \vspace*{0.3em}
    \LeftComment{\textbf{Setup Phase}}
    \For{$col \gets 0$ to $C - 1$} \label{lst:line:setup0}
    \State $M[0][C-1-col] \gets H_{\rho}\texttt{.squeeze}(b)$
    \EndFor \label{lst:line:setup0end}

    \For{$col \gets 0$ to $C - 1$}  \label{lst:line:setup1}
    \State $M[1][C-1-col] \gets M[0][col] \oplus H_\rho.\texttt{duplex}(M[0][col],b)$
    \EndFor \label{lst:line:setup1end}

    \For{$row^0 \gets 2$ to $R-1$} \label{lst:line:setup2}
    \State $prev^0 \gets row^0 - 1$
    \State $row^1 \gets row^0 - 2$
    \For{$col \gets 0$ to $C - 1$}
    \State $rand \gets H_\rho.\texttt{duplex}(M[row^1][col] \boxplus M[prev^0][col],b)$
    \State $M[row^0][C-1-col] \gets M[prev^0][col] \oplus rand$
    \State $M[row^1][col] \gets M[row^1][col] \oplus (rand \lll \omega)$
    \EndFor
    \EndFor \label{lst:line:setup2end}

    \vspace*{0.3em}
    \LeftComment{\textbf{Wandering Phase}}
    \For{$row^0 \gets 0$ to $R \cdot T-1$} \label{lst:line:wander}
    \State $prev^0 \gets row^0 - 1$

    \State $row^1 \gets \texttt{lsw}(rand) \texttt{ mod } R$ \label{lst:line:rowselection}
    \For{$col \gets 0$ to $C - 1$}
    \fontsize{8.75pt}{3.0pt}\selectfont\State $rand \gets H_\rho.\texttt{duplex}(M[row^1][col] \boxplus M[prev^0][col],b)$\label{lst:line:simplifiedadd}
    \small
    \State $M[row^0][col] \gets M[row^0][col] \oplus rand$ \label{lst:line:collide1}
    \State $M[row^1][col] \gets M[row^1][col] \oplus (rand \lll \omega)$ \label{lst:line:collide2}
    \EndFor
    \EndFor

    \vspace*{0.3em}
    \LeftComment{\textbf{Wrap-up Phase}} \label{lst:line:wrap}
    \State $H.\texttt{absorb}(M[row^1][0])$
    \State $K \gets H.\texttt{squeeze}(k)$

  \end{algorithmic}
\end{algorithm}

Lyra2 was initially created as a \gls{phs} for secure storage \cite{lyra2,lyra2TC}. Lyra2 uses the duplex construction from Keccak, where the permutation function $f$ is the round function from BLAKE2b. The reasoning for this choice is twofold and stems from the concept of favoring CPUs. On one hand, the G-function of BLAKE2b is software-oriented (e.g., the rotations are chosen to specifically benefit from SIMD instructions). On the other hand, the permutation of BLAKE2b has been shown to be secure even with a reduced number of rounds~\cite{Ji2009}, whereas a full permutation normally consists of 12 rounds. As explained in more detail in the sequel, after every permutation, the Lyra2 algorithm performs a memory access. A reduced number of rounds in a permutation allows more memory accesses for the same execution time, making low-memory attacks on parallel platforms more costly.

In the remainder of the text, calls to a full-round (i.e., $12$ iterations) duplex are denoted as calls to $H$, while reduced-round duplexing as calls to $H_\rho$, where $\rho$ denotes the reduced number of rounds. Because the G-functions are specified to operate on an array of $16$ $64$-bit words, Lyra2 uses a duplex with a width of $w = 16 \cdot 64 = 1024$\,bits. Pseudocode for the simplified version of Lyra2 that is used specifically in Lyra2REv2 is given in Algorithm\,\ref{algo:lyra2} and can be compared to the original Lyra2 pseudocode available in~\cite[Algorithm\,2]{lyra2}. In the following sections, we first explain each phase of the simplified Lyra2 algorithm used in Lyra2REv2 and how it differs from the reference implementation of Lyra2 in detail. Then, we explain the differences between Lyra2 used Lyra2REv2 and Lyra2MOD used in the updated Lyra2REv3 algorithm.

\subsection{Bootstrapping Phase}
In the bootstrapping phase, the duplex is initialized with a state that depends on the input $pwd$, a salt (which in Lyra2REv2 is set to be equal to $pwd$ for simplicity), and the parameters $T$, $R$, and $C$ by using a full-round absorb. The duplex $H$ in Algorithm\,\ref{algo:lyra2} internally uses a bitrate $b = 768$ bits and a capacity $c = 256$ bits. The $H.\texttt{absorb}(\cdot)$ call on line\,\ref{lst:line:boostrap}, however, considers only inputs of 512 bits instead of $b$ bits, so as to not overwrite the upper part of the initialization state, i.e, the 512-bit initialization value \textit{IV} specified by BLAKE2b. This results in two full-round absorbs, where the first and second absorbs process $(pwd \concat pwd)$ and \texttt{pad}$(params)$, respectively.

\subsection{Setup Phase}
During the setup phase of Lyra2, an $R \times C \times b$ memory matrix $M$ is initialized using the single-round duplex $H_{1}$. The simplified version of Lyra2 in Lyra2REv2 uses $R = C = 4$. Rows are initialized from first to last, while columns within each row are initialized from last to first. From the second row onward, a previous row is re-read, making it impractical to only store parts of the memory matrix. Also, from the third row onward, in addition to the previous row, i.e., $prev^0$, a specific pre-initialized row, i.e., $row^1$, is revisited (i.e., read and updated) in a deterministic manner. Rows are re-read or revisited from the first to the last column. Revisited rows use a rotated version of the duplex output, where the rotation number is chosen as $\omega = 64$ in Lyra2REv2. Note that the general revisiting scheme for $row^1$ is significantly more complicated when $R > 4$, as rows to be revisited can be chosen from within a specific window.

\subsection{Wandering Phase}\label{sec:SimpleLyra2:wandering}
The wandering phase is configurable to be the most time-consuming of the four phases. This is done through a timecost parameter $T$, that sets a number of rows $2R \cdot T$  to be revisited. In Lyra2REv2, there is only a single iteration over the memory matrix, as $T = 1$. Specifically, it revisits two rows $row^0$ and $row^1$, where $row^0$ is chosen deterministically but $row^1$ is chosen in a pseudorandom fashion by using the least significant part of the duplex output. Note that the pseudorandom and deterministic row can collide, resulting in the operations on lines~\ref{lst:line:collide1} and~\ref{lst:line:collide2} to sequentially read from and then write to the same matrix cell. Also note that the reference implementation of Lyra2 selects not only $row^1$, but also $row^0$ pseudorandomly. Furthermore, whereas the simplified Lyra2 in Lyra2REv2 uses a deterministic column counter $col$, the reference implementation features pseudorandom counters $col^0$ and $col^1$. Lastly, similar to $prev^0$ as the previous $row^0$, $prev^1$ is introduced to track the previous $row^1$. These extra variables appear, for example, on line~\ref{lst:line:simplifiedadd}, where the simplified Lyra2 has a two-operand word-wise addition, but the reference implementation would pass $M[row^0][col] \boxplus M[row^1][col] \boxplus M[prev^0][col^0] \boxplus M[prev^1][col^1]$ as input to the sponge.

\subsection{Wrap-up Phase}
The wrap-up phase consists of a full-round absorb of a specific cell of $M$ followed by a squeeze of the hashed output $K$. This specific cell is likewise pseudorandom, as it is selected as the first cell of the lastly revisited pseudorandom row. The requested squeeze length $k = 256$ is lower than the bitrate $b=768$, which means that the final output is provided directly from the duplex state without a permutation $f$.

\subsection{From Lyra2REv2 to Lyra2REv3}\label{sec:Lyra2REv3}
Recently, the developers of Lyra2REv2 proposed Lyra2REv3 with the goal to make ASIC miners for Lyra2REv2, that became available on the market, obsolete. Vertcoin is currently the only Lyra2REv2-based cryptocurrency that has performed a hard fork to force the miners to use Lyra2REv3~\cite{vertcoinfork}. Fig.~\ref{fig:Lyra2REv3} illustrates the new chained hashing algorithm. Compared to the Lyra2REv2 chain in Fig.~\ref{fig:Lyra2REv2}, it can be seen that the Keccak-256 and Skein-256 hashing algorithms were removed from the chain, and a second instance of a Lyra2-based hashing algorithm was added. The developers justified the removal of both Keccak-256 and Skein-256 by mentioning the existence of significantly more efficient hardware implementations of these algorithms compared to their software counterparts. In addition to these changes, the simplified Lyra2 algorithm itself has been modified.

\begin{algorithm}[t]
  \caption{Lyra2MOD algorithm as specified in Lyra2REv3.}
  \label{algo:lyra2mod}
  \begin{algorithmic}[1]\fontsize{8.75pt}{8.75pt}\selectfont
   	\State \texttt{PARAMS:} $H, \rho, \omega, T, R, C, k, b\text{ as }H.b, c \text{ as }H.c$
    \State\texttt{INPUT:} $pwd$
    \State \texttt{OUTPUT:} $K$

    \vspace*{0.3em}
    \LeftComment{\textbf{Bootstrapping Phase}}
    \textcolor{blue}{\State $instance \gets 0$} \label{lst:lyra2mod:initinstance}
    \State $params \gets \texttt{len}(K) \concat \texttt{len}(pwd) \concat \texttt{len}(pwd) \concat T \concat R \concat C$
    \State $H.\texttt{absorb}(\texttt{pad}(pwd \concat pwd \concat params))$

    \vspace*{0.3em}
    \LeftComment{\textbf{Setup Phase}}
    \For{$col \gets 0$ to $C - 1$}
    \State $M[0][C-1-col] \gets H_{\rho}\texttt{.squeeze}(b)$
    \EndFor

    \For{$col \gets 0$ to $C - 1$}
    \State $M[1][C-1-col] \gets M[0][col] \oplus H_\rho.\texttt{duplex}(M[0][col],b)$
    \EndFor

    \For{$row^0 \gets 2$ to $R-1$}
    \State $prev^0 \gets row^0 - 1$
    \State $row^1 \gets row^0 - 2$
    \For{$col \gets 0$ to $C - 1$}
    \State $rand \gets H_\rho.\texttt{duplex}(M[row^1][col] \boxplus M[prev^0][col],b)$
    \State $M[row^0][C-1-col] \gets M[prev^0][col] \oplus rand$
    \State $M[row^1][col] \gets M[row^1][col] \oplus (rand \lll \omega)$
    \EndFor
    \EndFor

    \vspace*{0.3em}
    \LeftComment{\textbf{Wandering Phase}}
    \For{$row^0 \gets 0$ to $R \cdot T-1$}
    \State $prev^0 \gets row^0 - 1$
  	\textcolor{blue}{\State $rand' \gets H_0.\texttt{squeeze'}(b+c)$} \label{lst:lyra2mod:row1start}
  	\textcolor{blue}{\State $instance \gets rand'[instance] \texttt{ mod } 16$}
  	\textcolor{blue}{\State $row^1 \gets rand'[instance] \texttt{ mod } R$} \label{lst:lyra2mod:row1stop}
    \For{$col \gets 0$ to $C - 1$}
    \fontsize{8.75pt}{3.0pt}\selectfont\State $rand \gets H_\rho.\texttt{duplex}(M[row^1][col] \boxplus M[prev^0][col],b)$
    \small
    \State $M[row^0][col] \gets M[row^0][col] \oplus rand$
    \State $M[row^1][col] \gets M[row^1][col] \oplus (rand \lll \omega)$
    \EndFor
    \EndFor

    \vspace*{0.3em}
    \LeftComment{\textbf{Wrap-up Phase}}
    \State $H.\texttt{absorb}(M[row^1][0])$
    \State $K \gets H.\texttt{squeeze}(k)$

  \end{algorithmic}
\end{algorithm}

The updated Lyra2 algorithm as used in Lyra2REv3, called \emph{Lyra2MOD}, is illustrated in Algorithm~\ref{algo:lyra2mod}, where the changes from the simplified Lyra2 algorithm used in Lyra2REv2 are highlighted in blue (lines \ref{lst:lyra2mod:initinstance}, and \ref{lst:lyra2mod:row1start}--\ref{lst:lyra2mod:row1stop}). While the changes appear to be minor, the Lyra2MOD modifications are non-conventional in the Lyra2 scheme. Lyra2MOD introduces a new variable called $instance$, that can take the value of the four least-significant bits of any word in the $(b+c)$-bit sponge state. This assignment is non-conventional, because it does not exclude the four words that make up the sponge capacity $c$. Within its specifications, the sponge construction does not allow for such an operation that directly reads bits from the capacity part of the sponge \cite{sponge}. The variable $instance$ is then used to update $row^1$, which can now similarly be assigned some least significant part of any state word. The assignments to $instance$ and $row^1$ require defining a new operation on the sponge $H$ that requests the current state without performing any rounds. We call this new operation $squeeze'$ for its similarity with the $squeeze$ operation, with the difference that the former is not restricted to requests of $l \leq b$ bits on the state. To omit the round functionality of the sponge, we call $squeeze'$ on $H_0$, i.e., the sponge reduced to zero rounds. The intended effect of these changes is to further serialize the algorithm, making hardware implementation more challenging. The impact of these changes on resource requirements and on performance is briefly described in Section~\ref{sec:impl:Lyra2MOD}.

\section{Programmable Logic Implementation of Simplified Lyra2}\label{sec:impl:SimpleLyra2}
This section describes how the Lyra2 algorithm, which is the most complex algorithm of the Lyra2REv2 chain, can be efficiently mapped to a hardware implementation. The hardware implementation of the full Lyra2REv2 hashing chain is discussed in Section~\ref{sec:impl:Lyra2REv2}, as well as the changes that would be required for a Lyra2REv3 chain. Similarly to the previous section, we first describe an implementation of Lyra2 for Lyra2REv2, and we then explain the necessary changes to implement Lyra2MOD for Lyra2REv3.

\begin{figure}[t]
  \centering
  \includegraphics[width=0.76\columnwidth]{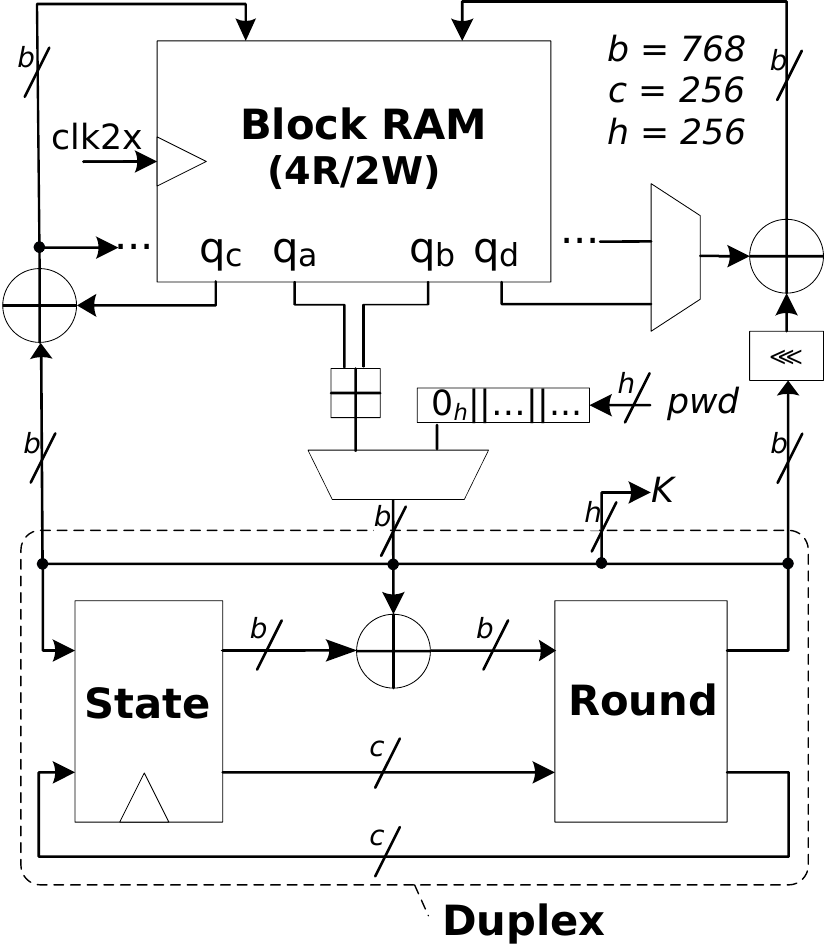}%
  \caption{Datapath of the Lyra2 PL architecture proposed in this work.}
  \label{fig:lyra2-data}%
\end{figure}

Recall that, in the current instance of Lyra2 as used in Lyra2REv2, the timecost parameter is $T=1$, the number of rows in the memory matrix is $R=4$, the number of columns in the memory matrix is $C=4$, and the desired hashing output length is $k=256$ (note that the same parameter values are also used for Lyra2MOD in Lyra2REv3). The architecture described in this work is optimized for these parameter values, but can be modified relatively easily to accommodate potential changes in the aforementioned parameters. Moreover, for $R=C=4$ and $b=768$, the memory matrix $M$ is 1.5~kB in size, which is clearly not prohibitively large to be implemented either in \gls{pl} or on an ASIC. The claimed ASIC-resistance of the Lyra2REv2 algorithm comes from the fact that $T$, $C$, and $R$ can be increased easily if necessary and that the chain of hashing algorithms itself can be modified (as is the case with the newer Lyra2REv3 algorithm).

The high-level datapath of the proposed \gls{pl} implementation of the simplified Lyra2 algorithm used in Lyra2REv2 is shown in Fig.~\ref{fig:lyra2-data}, where the duplex construction with its state, round, and XOR input block can be clearly distinguished. The memory matrix $M$ is mapped to a \gls{bram}. To reduce the complexity of the \gls{mux} at the input of the duplex, the \gls{bram} also contains constant vectors of $b$\,bits used during the bootstrapping and setup phases, i.e., an all-zero vector and the $\texttt{pad}(params)$ vector.

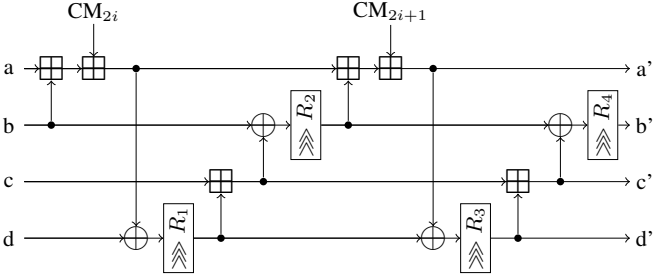
\begin{figure}[t]
  \resizebox{\columnwidth}{!}{%
    \begin{tikzpicture}[font=\small,inner sep=1pt, minimum width=1.2em]
      \def\nsep{0.6}
      \def\hsep{0.8}
      \def\vsep{15*\nsep}
      \def\os{-0.17}

      \node (a) at (0,0) {a};
      \node (b) at ($(a)+(0,-\hsep)$) {b};
      \node (c) at ($(b)+(0,-\hsep)$) {c};
      \node (d) at ($(c)+(0,-\hsep)$) {d};
      \node (cm1) at ($(a)+(2*\nsep,\hsep)$) {CM$_{2i}$};
      \node (cm2) at ($(a)+(9*\nsep,\hsep)$) {CM$_{2i+1}$};

      \node (ab1) at ($(a)+(\nsep,0)$) {\Large $\boxplus$};
      \node (ab2) at ($(a)+(2*\nsep,0)$) {\Large $\boxplus$};
      \node[branch] (abr1) at ($(a)+(3*\nsep,0)$) {};
      \node (ab3) at ($(a)+(8*\nsep,0)$) {\Large $\boxplus$};
      \node (ab4) at ($(a)+(9*\nsep,0)$) {\Large $\boxplus$};
      \node[branch] (abr2) at ($(a)+(10*\nsep,0)$) {};
      \node[branch] (bbr1) at ($(b)+(1*\nsep,0)$) {};
      \node (bo1) at ($(b)+(6*\nsep,0)$) {\Large $\oplus$};
      \node[draw=black] (bs1) at ($(b)+(7*\nsep,0)$) {\rotatebox{90}{$\ggg R_2$}};
      \node[branch] (bbr2) at ($(b)+(8*\nsep,0)$) {};
      \node (bo2) at ($(b)+(13*\nsep,0)$) {\Large $\oplus$};
      \node[draw=black] (bs2) at ($(b)+(14*\nsep,0)$) {\rotatebox{90}{$\ggg R_4$}};
      \node (cb1) at ($(c)+(5*\nsep,0)$) {\Large $\boxplus$};
      \node[branch] (cbr1) at ($(c)+(6*\nsep,0)$) {};
      \node (cb2) at ($(c)+(12*\nsep,0)$) {\Large $\boxplus$};
      \node[branch] (cbr2) at ($(c)+(13*\nsep,0)$) {};
      \node (do1) at ($(d)+(3*\nsep,0)$) {\Large $\oplus$};
      \node[draw=black] (ds1) at ($(d)+(4*\nsep,0)$) {\rotatebox{90}{$\ggg R_1$}};
      \node[branch] (dbr1) at ($(d)+(5*\nsep,0)$) {};
      \node (do2) at ($(d)+(10*\nsep,0)$) {\Large $\oplus$};
      \node[draw=black] (ds2) at ($(d)+(11*\nsep,0)$) {\rotatebox{90}{$\ggg R_3$}};
      \node[branch] (dbr2) at ($(d)+(12*\nsep,0)$) {};
      
      \node (ap) at ($(a)+(\vsep,0)$) {a'};
      \node (bp) at ($(b)+(\vsep,0)$) {b'};
      \node (cp) at ($(c)+(\vsep,0)$) {c'};
      \node (dp) at ($(d)+(\vsep,0)$) {d'};

      \draw[->] (a) edge ($(ab1)+(\os,0)$) edge ($(ab2)+(\os,0)$) edge ($(ab3)+(\os,0)$) edge ($(ab4)+(\os,0)$) edge (ap);
      \draw[->] (b) edge ($(bo1)+(\os,0)$) edge (bs1);
      \draw[->] (bs1) edge ($(bo2)+(\os,0)$) edge (bs2);
      \draw[->] (bs2) edge (bp);
      \draw[->] (c) edge ($(cb1)+(\os,0)$) edge ($(cb2)+(\os,0)$) edge (cp);
      \draw[->] (d) edge ($(do1)+(\os,0)$) edge (ds1);
      \draw[->] (ds1) edge ($(do2)+(\os,0)$) edge (ds2);
      \draw[->] (ds2) edge (dp);

      \draw[->] (bbr1) -- ($(ab1)+(0,\os)$);
      \draw[->] (cm1) -- ($(ab2)+(0,-\os)$);
      \draw[->] (abr1) -- ($(do1)+(0,-\os)$);
      \draw[->] (dbr1) -- ($(cb1)+(0,\os)$);
      \draw[->] (cbr1) -- ($(bo1)+(0,\os)$);
      \draw[->] (bbr2) -- ($(ab3)+(0,\os)$);
      \draw[->] (cm2) -- ($(ab4)+(0,-\os)$);
      \draw[->] (abr2) -- ($(do2)+(0,-\os)$);
      \draw[->] (dbr2) -- ($(cb2)+(0,\os)$);
      \draw[->] (cbr2) -- ($(bo2)+(0,\os)$);

    \end{tikzpicture}%
    }%
  \caption{Hardware architecture of the BLAKE G-function (adapted from \cite[Fig.\,2.1]{Aumasson2008}). All signals are $m$ bits wide.}
  \label{fig:blake}%
\end{figure}

As mentioned in Sections~\ref{sec:bg:blake2b} and \ref{sec:SimpleLyra2}, the round function $f$ of the Lyra2 algorithm is an arrangement of BLAKE G-functions. Fig.~\ref{fig:blake} shows the hardware architecture of BLAKE's G-function, where all signals are $m$ bits wide. Lyra2 uses the BLAKE2b variation, i.e., $m=64$, $R_1 = 32$, $R_2 = 24$, $R_3 = 16$, and $R_4 = 63$ (cf. Algorithm\,\ref{algo:blake2b:g}). Furthermore, the CM$_{2i}$ and CM$_{2i+1}$ inputs are not used, thus the corresponding adders are omitted in the implementation of the round function for Lyra2 presented in this work.

In the following, we first describe a version of the hardware architecture of the simplified Lyra2 core described in this work, where each round of the $f$ function is executed in a single \gls{cc}. We then describe how this basic architecture can be improved through pipelining.

\subsection{Basic Iterative Architecture}

The basic iterative Lyra2 architecture requires 68 \glspl{cc} per hash: 24 for the bootstrapping phase, 16 each for the setup and wandering phases, and 12 for the wrap-up phase.

\subsubsection{Bootstrapping Phase} During the bootstrapping phase, the duplex processes two 512-bit input blocks from $\texttt{pad}(pwd \concat pwd \concat params)$ using a full-round absorb. In Lyra2REv2, $pwd = cube_{\text{out}}$, with $cube_{\text{out}}$ being the output of the first CubeHash instance, i.e., the previous algorithm in the chain. Thus, as shown in Fig.\,\ref{fig:lyra2-data}, the $(pwd \concat pwd)$ vector is one of the inputs to the \gls{mux} of the duplex. On the other hand, the $\texttt{pad}(params)$ vector is fed into the sponge by loading it on $q_a$ while simultaneously loading the all-zero vector on $q_b$. Both constants are stored at known addresses in the \gls{bram}, and are absorbed in a separate 12-round \texttt{Bootstrap} state. During bootstrapping, the duplex only receives an input vector in the first round. Hence, for subsequent rounds, $q_a$ and $q_b$ output the all-zero vector, and their sum is passed to the duplex via its input \gls{mux}.

\subsubsection{Setup Phase} The setup phase is split into three distinct phases for convenience, namely \texttt{Setup0}, \texttt{Setup1}, and \texttt{Setup2}, which correspond to Lines~\ref{lst:line:setup0}--\ref{lst:line:setup0end}, Lines~\ref{lst:line:setup1}--\ref{lst:line:setup1end}, and Lines~\ref{lst:line:setup2}--\ref{lst:line:setup2end} of Algorithm~\ref{algo:lyra2}, respectively. Similarly to the bootstrapping phase, the setup phase uses the all-zero vector stored in the \gls{bram}. In the \texttt{Setup0} state, the squeezes input an empty message into the duplex and directly write the duplex output to the \gls{bram}. To achieve that, the all-zero vector is output on $q_a$, $q_b$, and $q_c$. \texttt{Setup1} reads the all-zero vector on $q_b$, but a specific vector from the \gls{bram} on $q_a$. \texttt{Setup2} reads two vectors from $q_a$ and $q_b$. Both the duplex output and the rotated duplex output are XOR'd with two other vectors from the \gls{bram}, requiring the two XOR blocks in parallel as illustrated in Fig.\,\ref{fig:lyra2-data}. On the control path, counters keep track of the various rows ($row^0, row^1, prev^1$) and their corresponding columns to generate read and write addresses for the RAM.

\subsubsection{Wandering Phase} The input to the duplex in the wandering phase is always the word-wise addition of two RAM cells. Both XOR blocks connected to the duplex output are used. As mentioned in the algorithmic description of the wandering phase in Section~\ref{sec:SimpleLyra2:wandering}, the pseudorandom and deterministic rows used in this phase can collide. In hardware, this special case requires the output of one XOR block to be input to the other, while the write port of the first XOR block needs to be disabled to prevent write collisions on the RAM.

\subsubsection{Wrap-Up Phase} During the wrap-up phase, one RAM cell is input into the sponge and then processed using a full-round absorb. For the following squeeze, the requested hashed-output length $k$ is lower than the bitrate $b$, i.e., the duplex state directly provides the output hash.

\subsection{Memory Matrix}
In the wandering phase, up to two RAM cells need to be written and three RAM cells need to be read per \gls{cc}. These operations cannot be spread over multiple \glspl{cc} without negatively affecting the overall throughput of the design. Therefore, we use standard true-dual-port \glspl{bram} along with multipumping and replication techniques~\cite{Laforest2010} in order to implement the required functionality. Replication provides extra read ports by physically replicating the \gls{bram} while connecting the write ports to keep the two copies coherent. Multipumping operates the \gls{bram} at double the clock frequency of the surrounding logic, which, together with replication, effectively provides four read ports and two write ports. A $b=768$-bit wide \gls{bram} with true-dual-port functionality can be implemented using $21 \times 36K$ and one $18K$ \gls{pl} BRAM primitives, which are $21 \times 36$ and $18$ bits wide, respectively. 
In total, the Lyra2 core then uses $42 \times 36K + 2 \times 18K = 1548$ Kbits of \gls{bram}.

\subsection{Pipelined Architecture}\label{sec:impl:SimpleLyra2:Pipelined}

Pipelining the BLAKE2b round function can greatly reduce the delay of the critical path. Recall that the round function consists of an arrangement of G-functions, whose architecture is illustrated in Fig.~\ref{fig:blake}. In the basic iterative version described above, the critical path extends from the RAM read ports to the RAM write ports and contains eight sequential 64-bit adders in the round function. Dividing these sequential adders into eight pipeline stages greatly increases the achievable clock frequency, with only a minimal increase to resource usage due to the additional registers required. Each hash that is concurrently being processed in the pipeline needs its own memory. However, extra RAM-based memory is readily available since the current Lyra2REv2 parameters result in a RAM depth much shallower than that of the \gls{pl} \glspl{bram}. With adequate scheduling, concurrent hashes write to the same \glspl{bram} in distinct \glspl{cc}. While read ports $q_a$ and $q_b$ feed the duplex, $q_c$ and $q_d$ feed the XORs with duplex outputs. When pipelining the round function, $q_c$ and $q_d$ therefore need to be delayed by as many \glspl{cc} as there are pipeline stages. The extra read port that is unused in the basic architecture allows delaying the control path for $q_d$ rather than using a delayed version of $q_b$, avoiding a long chain of 768-bit registers. Eight pipeline stages in the round increase the latency to 544 \glspl{cc} per hash. On the other hand, the pipeline can process eight hashes concurrently, i.e., one hash is output every 68 \glspl{cc} on average. Finally, the logic depth reduction, from eight sequential 64-bit adders to a single one, more than doubles the achievable clock frequency, which in turn significantly increases the overall hashing throughput of the pipelined architecture.

\subsection{Programmable Logic Implementation of Lyra2MOD}\label{sec:impl:Lyra2MOD}
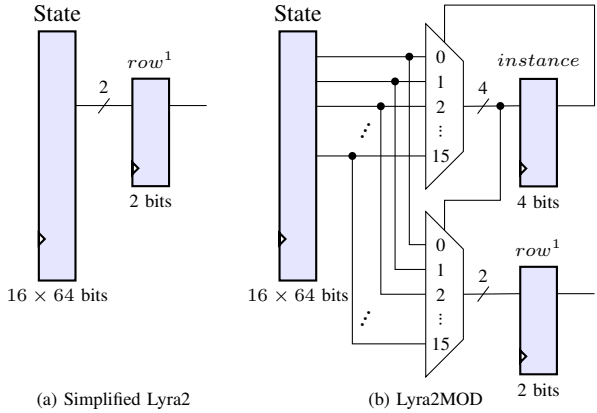
\begin{figure}[t]
  \centering
  \tikzstyle{branch}=[fill,shape=circle,minimum size=3pt,inner sep=0pt]
  \tikzstyle{inv}=[draw,circle,xshift=-0.62mm,fill=white,minimum size=3.5pt,inner sep=0pt]

  \begin{tikzpicture}[font=\scriptsize,inner sep=1pt, minimum width=1.2em]
    \node[shape=mux5invsel] (m0) at (2.0,0) {};
    \node[shape=mux5invsel] (m1) at ($(m0.south)+(0,-1.4)$) {};

    \node[shape=reg5] (state) at ($(m0.in4)+(-1.7,0)$) {};
    \node (statelabel) at ($(state.north)+(0,.25)$) {\small State};
    \node (statelabel2) at ($(state.south)-(0,.2)$) {$16 \times 64$ bits};

    \node[shape=reg] (idx) at ($(m0.out)+(1.0,-0.35)$) {};
    \node (idxlabel) at ($(idx.north)+(0,.25)$) {$instance$};
    \node (idxlabel2) at ($(idx.south)-(0,.2)$) {4 bits};
    \node[shape=reg] (row) at ($(m1.out)+(1.0,-0.35)$) {};
    \node (rowlabel) at ($(row.north)+(0,.25)$) {$row^1$};
    \node (rowlabel2) at ($(row.south)-(0,.2)$) {2 bits};

    \node at ($(m0.in0)+(0.2,0)$) {0};
    \node at ($(m0.in1)+(0.2,0)$) {1};
    \node at ($(m0.in2)+(0.2,0)$) {2};
    \node at ($(m0.in3)+(0.2,0)$) {\rotatebox{90}{...}};
    \node at ($(m0.in4)+(0.2,0)$) {15};

    \node at ($(m1.in0)+(0.2,0)$) {0};
    \node at ($(m1.in1)+(0.2,0)$) {1};
    \node at ($(m1.in2)+(0.2,0)$) {2};
    \node at ($(m1.in3)+(0.2,0)$) {\rotatebox{90}{...}};
    \node at ($(m1.in4)+(0.2,0)$) {15};

    \draw[-] (m0.sel) -- ($(m0.sel)+(0,.5)$) -- ($(m0.sel -| idx.Q) + (.5,.5)$) -- ($(idx.Q) + (.5,0)$) -- (idx.Q);
    \draw[-] (m0.out) -- (idx.D);

    \draw[-] ($(m0.out)!0.35!(idx.D)-(.075,.1)$) -- ($(m0.out)!0.35!(idx.D)+(.075,.1)$);
    \node at ($(m0.out)!0.35!(idx.D)+(0,.25)$) {4};

    \draw[-] ($(m0.out)!0.65!(idx.D)$) node[branch] {} |- ($(m0.south)!0.5!(m1.north)$) -- (m1.sel);

    \draw[-] (m1.out) -- (row.D);
    \draw[-] ($(m1.out)!0.35!(row.D)-(.075,.1)$) -- ($(m1.out)!0.35!(row.D)+(.075,.1)$);
    \node at ($(m1.out)!0.35!(row.D)+(0,.25)$) {2};

    \draw[-] (state.Q0) |- (m0.in0);
    \draw[-] (state.Q1) |- (m0.in1);
    \draw[-] (state.Q2) |- (m0.in2);
    \draw[-] (state.Q4) |- (m0.in4);

    \draw[-] ($(state.Q0)!.85!(m0.in0)$) node[branch] {} |- (m1.in0);
    \draw[-] ($(state.Q1)!.72!(m0.in1)$) node[branch] {} |- (m1.in1);
    \draw[-] ($(state.Q2)!.59!(m0.in2)$) node[branch] {} |- (m1.in2);
    \draw[-] ($(state.Q4)!.33!(m0.in4)$) node[branch] {} |- (m1.in4);

    \node (p1) at ($(state.Q4)!.33!(m0.in4)$) {};
    \node (p2) at ($(state.Q2)!.59!(m0.in2)$) {};

    \draw[decoration={text along path,text={...},text align={center}},decorate] (p1) -- (p2);

    \draw[decoration={text along path,text={...},text align={center}},decorate] (p1 |- m1.in4) -- (p2 |- m1.in2);

    \draw[-] (row.Q) -- ($(row.Q)+(.5,0)$);

    \node[shape=reg5] (state2) at ($(state)+(-3.2,0)$) {};
    \node (state2label) at ($(state2.north)+(0,.25)$) {\small State};
    \node (state2label2) at ($(state2.south)-(0,.2)$) {$16 \times 64$ bits};
    \node[shape=reg] (row2) at ($(state2.Q2)+(1.0,-0.35)$) {};
    \node (row2label) at ($(row2.north)+(0,.25)$) {$row^1$};
    \node (row2label2) at ($(row2.south)-(0,.2)$) {2 bits};

    \draw[-] (state2.Q2) |- (row2.D);
    \draw[-] (row2.Q) -- ($(row2.Q)+(.5,0)$);
    \draw[-] ($(state2.Q2)!0.5!(row2.D)-(.075,.1)$) -- ($(state2.Q2)!0.5!(row2.D)+(.075,.1)$);
    \node at ($(state2.Q2)!0.5!(row2.D)+(0,.25)$) {2};

    \node at ($(row.south -| row2.west)+(-0.25,-0.35)$) {(a) Simplified Lyra2};
    \node at ($(row.south -| m1.west)+(0,-0.35)$) {(b) Lyra2MOD};

  \end{tikzpicture}
  \caption{Hardware implementation of the row selection during the wandering phase for (a) the simplified Lyra2 and (b) Lyra2MOD. Clock signals are omitted for clarity.}
  \label{fig:SimpleLyra2-vs-Lyra2MOD}
\end{figure}

A \gls{pl} implementation of the Lyra2MOD algorithm can be based on the pipelined architecture of the simplified Lyra2 algorithm as described in Section~\ref{sec:impl:SimpleLyra2:Pipelined}, with appropriate changes to support the modified wandering phase explained in Section~\ref{sec:Lyra2REv3}. Fig.~\ref{fig:SimpleLyra2-vs-Lyra2MOD} shows the hardware implementation of the row selection during the wandering phase for both the simplified Lyra2 (Lyra2REv2) and Lyra2MOD (Lyra2REv3) algorithms. Specifically, Fig.~\ref{fig:SimpleLyra2-vs-Lyra2MOD}(a) shows that in the simplified Lyra2 algorithm, the row is selected simply based on the $2$ least-significant bits of the state (cf. line~\ref{lst:line:rowselection} of Algorithm\,\ref{algo:lyra2}).
On the other hand, the row selection in Lyra2MOD is much more involved (cf. lines~\ref{lst:lyra2mod:row1start}--\ref{lst:lyra2mod:row1stop} of Algorithm\,\ref{algo:lyra2}). Thus, as shown in Fig.~\ref{fig:SimpleLyra2-vs-Lyra2MOD}(b), Lyra2MOD requires the addition of multiplexers and memory to store the new $instance$ variable. The $instance$-variable memory is initialized to all zeros during the bootstrap phase. Finally, note that in the $8$-stage pipelined architecture, $instance$ and $row^1$ need to be stored for every hash in the pipeline using small $8 \times 4$\,bits and $8 \times 2$\,bits RAMs, respectively.  After implementation, verification, and synthesis of Lyra2MOD, it was found that the changes introduced have negligible impact in terms of resources. Furthermore, the critical path is unaffected as the new row-selection logic in Lyra2MOD translates to significantly fewer logic levels than that of the $64$-bit adders on the datapath.

\section{\glsentryshort{mpsoc} Implementation of a Standalone Lyra2REv2 Miner}\label{sec:impl:Lyra2REv2}
In this section, we present an \gls{mpsoc}-based architecture for the standalone Lyra2REv2 miner, and the changes that would be required to support the Lyra2REv3 chain. Specifically, we implement the computation-intensive part of the Lyra2REv2 (or Lyra2REv3) chained hashing algorithm on the \gls{pl} along with supporting logic, and use the \gls{ps} capabilities of the \gls{mpsoc} to run supporting software that is used to handle high-level cryptocurrency protocol tasks.
\begin{figure*}[t]
  \centering
  \resizebox{0.9\textwidth}{!}{
    \begin{tikzpicture}[%
      font=\scriptsize,%
      inner sep=0pt,%
      minimum width=1.0em,%
      square/.style={regular polygon,regular polygon sides=4},%
      ]
      \tikzset{multiple/.style = {%
          double copy shadow={shadow xshift=.3em,shadow yshift=-.25em,draw=black!60},%
          fill=white,draw=black,thick,minimum height = 2.2em,minimum width=3.1em},%
        ordinary/.style = {rectangle,draw,thick,minimum height = 1cm,minimum width=2cm}}

      \def\nsep{0.65}
      \def\fifowidth{1.75em}
      \def\schedwidth{1.2em}

      \node[draw=black,minimum height=0.95cm,inner sep=2pt] (cmm) at (0,0) {\shortstack{customized\\cpuminer-\\[-2pt]multi}};
      \node[draw=black,minimum height=0.95cm,inner sep=2pt,anchor=west] (driver) at ($(cmm.east)+(0.75,0)$) {\shortstack{FPGA-miner\\driver}};
      \node[draw=black,minimum height=0.95cm,inner sep=2pt,anchor=west] (kernel) at ($(driver.east)+(0,0)$) {\rotatebox{90}{kernel}};

      \draw[-,dashed] ($(cmm.east)!0.5!(driver.west)+(0,.85)$) -- ($(cmm.east)!0.5!(driver.west)-(0,.60)$);
      \draw[<->] (cmm.east) -- node[above,yshift=1pt,fill=white,inner ysep=2pt] {\tiny MMAP} (driver.west);
      \node[anchor=east] (userspace) at ($(cmm.north east)+(0.2,0.3)$) {User Space};
      \node[anchor=west] (kernelspace) at ($(driver.north west)+(-0.2,0.3)$) {Kernel Space};
      \node[anchor=south] (linux) at ($(userspace.north east)!0.5!(kernelspace.north west)+(0,0.2)$) {\bf GNU/Linux};
      
      \node (linux) [draw=black, inner ysep=0.3em, inner xsep=0.2em, xshift=-0.2em, fit = (linux) (cmm) (kernel) (userspace)] {};
      
      \node[anchor=south] (PS) at ([yshift=.8em]linux.north) {\small\bf Processing System (PS)};

      \node[draw=black,minimum height=7.8em,inner sep=2pt] (fifo-if) at ($(kernel.east)+(1.25,0)$) {\tiny\shortstack{Memory-\\Mapped\\ Adapter}};
      \node[anchor=north,yshift=-.5em] at (fifo-if.north) {\shortstack{Reg.\\File}};

      \node[anchor=west] (PL) at (PS -| fifo-if.west) {\small\bf Programmable Logic (PL)};
      \draw[-,dashed] ($(kernel.east |- PS.east)!0.5!(PL.west)+(0,.1)$) -- ($(kernel.east |- linux.south)!0.5!(PL.west |- fifo-if.south)-(0,.2)$);
      
      \draw[<->] (kernel) -- node[above,yshift=0em,fill=white,inner ysep=6pt] {\shortstack{AXI4-\\Lite}} node[below,yshift=-0em,fill=white,inner ysep=6pt] {\textcolor{blue}{\tiny 32}} node[below,yshift=0.25em] {\textcolor{blue}{\scriptsize /}} (fifo-if);

      \node[draw=black,minimum height=3.8em, minimum width=\fifowidth, anchor=north west,inner sep=2pt] (infsm) at ($(fifo-if.north east)+(\nsep,0)$) {\shortstack{Input\\Control\\FSM}};
      \draw[<->] ([yshift=0.6em]infsm.west) -- node[above, inner ysep=2pt] {\tiny \texttt{ctrl}} ++(left:\nsep);
      \draw[<-] ([yshift=-0.6em]infsm.west) -- node[above, inner ysep=2pt] {\tiny \texttt{newbl}} node[below, inner ysep=2pt, yshift=1pt] {\tiny \texttt{data}} ++(left:\nsep);

      \node[draw=black,minimum width=1.65cm, anchor=north west, inner sep=2pt] (chain) at ($(infsm.north east)+(\nsep,0)$) {\shortstack{Lyra2REv2\\Chain}};
      \draw[<->] ([yshift=0.5em]chain.west) -- node[above, inner ysep=2pt] {\tiny \texttt{ctrl}} ++(left:\nsep);
      \draw[<-] ([yshift=-0.5em]chain.west) -- node[above, inner ysep=2pt] {\tiny \texttt{block}} ++(left:\nsep);

      \node[draw=black,minimum width=1.65cm, anchor=south west, inner sep=2pt] (metafifo) at ($(infsm.south east)+(\nsep,0)$) {\shortstack{Metadata\\FIFO}};
      \draw[<-] (metafifo.west) -- node[above, inner ysep=2pt] {\tiny \texttt{meta}} ++(left:\nsep);

      \node[draw=black,anchor=north west, inner sep=2pt, minimum width=\fifowidth, minimum height=1.1
      25cm] (thresverif) at ($(chain.north east)+(\nsep,0)$) {\shortstack{Thres.\\Verif.}};
      \draw[->] (chain.east) -- node[above, inner ysep=2pt] {\tiny \texttt{hash}} ++(right:\nsep);
      \draw[->] ([yshift=0.5em]metafifo.east) -- node[above, inner ysep=2pt] {\tiny \texttt{thres}} ++(right:\nsep);

      \node[draw=black,minimum height=3.0em, minimum width=\fifowidth, anchor=south west,inner sep=2pt] (outfsm) at ($(fifo-if.south east)+(\nsep,0)$) {\shortstack{Output\\Control\\FSM}};
      \draw[<->] ([yshift=0.6em]outfsm.west) -- node[above, inner ysep=2pt] {\tiny \texttt{ctrl}} ++(left:\nsep);
      \draw[->] ([yshift=-0.6em]outfsm.west) -- node[above, inner ysep=3pt] {\tiny \texttt{result}} ++(left:\nsep);

      \draw[->] (infsm.south) -- node[right, xshift=1pt] {\tiny \texttt{flush}} (outfsm.north);
      \node (pos1) at ($([yshift=-0.5em]metafifo.east)+(0.5*\nsep,0)$) {};
      \node (pos2) at ([yshift=0.5em]outfsm.east) {};
      \draw[->] ([yshift=-0.5em]metafifo.east) -- (pos1.center) -| (pos1.center |- pos2.center) -- node[above, inner ysep=2pt] {\tiny \texttt{nonce}} ([yshift=0.5em]outfsm.east);

      \node (pos1) at ($(thresverif.east)+(0.5*\nsep,0)$) {};
      \node (pos2) at ([yshift=-0.5em]outfsm.east) {};
      \draw[->] (thresverif.east) -- ++(right:0.5*\nsep) |- (pos1.center) -- (pos1.center |- pos2.center) -- node[above, inner ysep=2pt, xshift=-2em] {\tiny \texttt{success}} ([yshift=-0.5em]outfsm.east);
    \end{tikzpicture}
  }%
  \caption{Architecture of the \glsentryshort{mpsoc} implementation of the standalone Lyra2REv2 miner. The computation-intensive Lyra2REv2 chained hashing algorithm is implemented on the \gls{pl} of the device along with nonce generation with threshold verification. Software that is used to handle high-level cryptocurrency protocol tasks runs on the \gls{ps} of the device.}
  \label{fig:Lyra2REv2-arch}
\end{figure*}
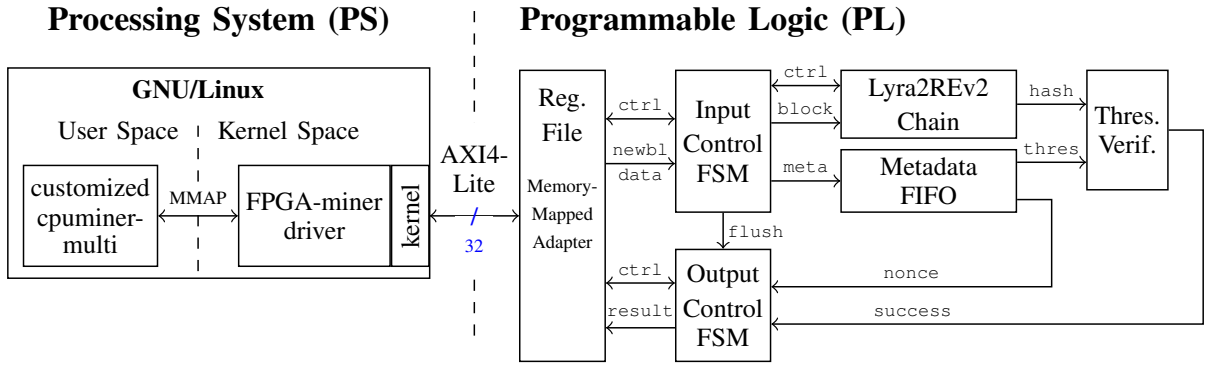

Fig.~\ref{fig:Lyra2REv2-arch} shows the high-level architecture of the proposed standalone Lyra2REv2 miner, where the supporting software on the \gls{ps} side (left) provides the \gls{pl} side (right) of the miner with the required data to start the search for a nonce that leads to a hash that meets the target threshold. In case of success, the supporting software reads back the winning nonce from the register file, regenerates the winning hash, and communicates the results to the network through the high-level cryptocurrency protocol.

In the following, the miner is described in more details. We first briefly describe the communication mechanism between the \gls{ps} and the \gls{pl} sides of the \gls{mpsoc}. Then, we discuss the software on the \gls{ps}. The next two sections describe the control logic, including nonce generation and threshold verification. Lastly, we discuss the hardware implementation of the mining algorithm on the \gls{pl} side of the device including the hashing algorithms, other than simplified Lyra2 and Lyra2MOD that we have already described above.

\subsection{Communications Between the Processing System and the Programmable Logic}
\begin{figure}
  \centering
  \begin{tikzpicture}[%
    font=\scriptsize,%
    inner sep=0pt,%
    ]
    \def\nsep{0.2}
    \draw (0, 0) rectangle (32*\nsep, -0.4);
    \node at (8*\nsep, -0.2) {Version};
    \node at (23*\nsep, -0.2) {Reserved};
    \draw[-] (16*\nsep, 0) -- (16*\nsep, -0.4);
    \draw[-] (29*\nsep, 0) -- node[right,xshift=1pt]{\tiny E} (29*\nsep, -0.4);
    \draw[-] (30*\nsep, 0) -- node[right,xshift=0.5pt]{\tiny W} (30*\nsep, -0.4);
    \draw[-] (31*\nsep, 0) -- node[right,xshift=1pt]{\tiny N} (31*\nsep, -0.4);

    \draw[-] (0, 0.2) -- node[right,xshift=1pt]{\tiny 31} (0, 0);
    \draw[-] (16*\nsep, 0.1) -- node[left,xshift=-1pt,yshift=1pt,anchor=east]{\tiny 16} node[right,xshift=1pt,yshift=1pt]{\tiny 15} (16*\nsep, 0);
    \draw[-] (29*\nsep, 0.1) -- node[right,xshift=1pt,yshift=1pt]{\tiny 2} (29*\nsep, 0);
    \draw[-] (30*\nsep, 0.1) -- node[right,xshift=1pt,yshift=1pt]{\tiny 1} (30*\nsep, 0);
    \draw[-] (31*\nsep, 0.1) -- node[right,xshift=1pt,yshift=1pt]{\tiny 0} (31*\nsep, 0);
    \draw[-] (32*\nsep, 0.2) -- (32*\nsep, 0);

    \draw[->] (29.5*\nsep, -0.4) |- (27*\nsep, -0.6) node[anchor=east, inner sep=1pt] {Error};
    \draw[->] (30.5*\nsep, -0.4) |- (27*\nsep, -0.85) node[anchor=east, inner sep=1pt] {Winning nonce found};
    \draw[->] (31.5*\nsep, -0.4) |- (27*\nsep, -1.1) node[anchor=east, inner sep=1pt] {Nonce not found};

    \node at (16*\nsep, -1.5) {(a) Status Register (0x00)};

    \coordinate (S1) at (0, -2.0) {};
    \draw (S1) rectangle ($(S1)+(32*\nsep, -0.4)$);
    \node at ($(S1)+(15*\nsep, -0.2)$) {Reserved};
    \draw[-] ($(S1)+(31*\nsep, 0)$) -- node[right,xshift=1pt]{\tiny S} ($(S1)+(31*\nsep, -0.4)$);
    \draw[-] ($(S1)+(0, 0.2)$) -- node[right,xshift=1pt]{\tiny 31} (S1);
    \draw[-] ($(S1)+(31*\nsep, 0.1)$) -- node[left,xshift=-1pt,yshift=1pt]{\tiny 1} node[right,xshift=1pt,yshift=1pt]{\tiny 0} ($(S1)+(31*\nsep,0)$);
    \draw[-] ($(S1)+(32*\nsep, 0.2)$) -- ($(S1)+(32*\nsep,0)$);
    \draw[->] ($(S1)+(31.5*\nsep, -0.4)$) |- ($(S1)+(27*\nsep, -0.6)$) node[anchor=east, inner sep=1pt] {Start New Block};

    \node at (16*\nsep, -3.0) {(b) Control Register (0x04)};
  \end{tikzpicture}
  \caption{Detailed description of the (a) status and (b) control registers.}
  \label{fig:impl:registers}
\end{figure}
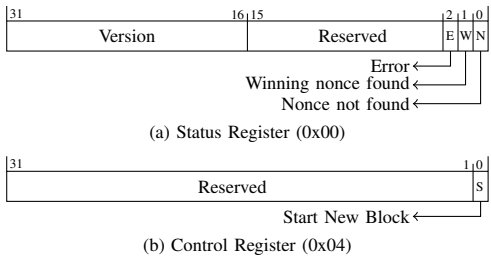

Given the limited amount of data that transits between the \gls{pl} and the \gls{ps}, a flip-flop based register file is used. Table~\ref{tab:impl:regfile} shows the content of the register file and Fig.~\ref{fig:impl:registers} provides a detailed view of the status and control registers. This allows the verification software to easily write $640$-bit block headers, $256$-bit target thresholds, and maximum nonces to the hardware miner, and to read back $32$-bit nonces, while reading and writing status and control signals. As shown in Fig.~\ref{fig:Lyra2REv2-arch}, the register file is wrapped in an adapter to allow access through a memory-mapped 32-bit wide AXI4-Lite bus, which is clocked at $250$\,MHz.

\begin{table}
  \centering
  \caption{Register file where the addresses are in bytes and each location holds 32 bits.}
  {\begin{tabular}{c|c}
    \toprule\hline
    \textbf{Addr}		& \textbf{Register} \\
    \hline\hline
    0x00			& Status \\\hline
    0x04		& Control \\\hline
    0x08		& Winning Nonce \\\hline
    0x0C			& \multirow{3}{*}{\vspace*{-5pt}\shortstack{Target\\Threshold}} \\
    \rotatebox{90}{$\cdots$} &  \\
    0x28			&  \\\hline
    0x2C			& \multirow{3}{*}{\vspace*{-5pt}\shortstack{Block\\Header}} \\
    \rotatebox{90}{$\cdots$} &  \\
    0x78			&  \\\hline
    0x7C			& Maximum Nonce \\
    \hline\bottomrule
  \end{tabular}}
  \label{tab:impl:regfile}
\end{table}

\subsection{Software on the Processing System}
The verification software consists of a Linux driver and a userspace application running inside a custom embedded GNU/Linux distribution. The driver exposes the memory-mapped register file as an $mmap()$ capable character device. Userspace applications can then use the character device to write block headers and interact with the \gls{pl} side of the miner.

The userspace application is based on the existing cpuminer-multi~\cite{cpuminer} open-source mining software, which was enhanced by adding a new type of algorithm, namely lyra2rev-hw. This new algorithm communicates directly with the mining hardware on the \gls{pl} side using the character device mentioned above. It writes the block header---which includes the starting nonce value---, the target threshold, and a maximum nonce value into the register file. Then, it asserts a bit in the control register to signal that new block data is available and starts to poll bits in the status register until either the \texttt{winning-nonce-found} bit is set or until the \texttt{nonce-not-found} bit is set. In the first case, the winning nonce is read back from the register file, the winning hash is regenerated and the results are communicated back to the network through the high-level cryptocurrency protocol. In the second case, the software proceeds with the next block header.

To ensure reliable and reproducible software builds, the Linux-based firmware and boot image are created using a customized Yocto~\cite{yocto} \gls{bsp}. This \gls{bsp} includes a custom layer on top of Xilinx's base Yocto \gls{bsp} and a set of supporting scripts to build and flash a boot image onto an SD card. The custom layer contains the patches to the Linux kernel and the patches to cpuminer-multi described above.

\subsection{Input Control Finite-State Machine and Nonce Generation in Programmable Logic}
The input control \gls{fsm} monitors the register file to detect when a new nonce search should be performed. Starting a new search implies stopping the on-going search by flushing the Lyra2REv2 chain pipeline. A new search begins by loading the block header, starting nonce, target threshold, and maximum nonce value from the register file into an internal memory. The block header is then fed to the Lyra2REv2 chain where the nonce is monotonically increased until either the maximum nonce value is reached or a new search is initiated. Meanwhile, the metadata FIFO is fed with the current nonce and target threshold.

\subsection{Threshold Verification and Output Finite-State Machine in Programmable Logic}
The threshold-verification logic reads the target threshold from the metadata FIFO and the hash output by the Lyra2REv2 chain, and uses a 256-bit comparator to determine whether the generated hash meets the threshold. As mentioned in Section \ref{sec:bg:pow}, for a \gls{pow} to be accepted by the network, the miner has to find a nonce that results in a hash with a value strictly smaller than the target threshold.
If the criterion is satisfied, the threshold-verification logic signals the output control \gls{fsm} that the winning nonce was found. The output control \gls{fsm} then reads the corresponding winning nonce from the metadata FIFO, asserts the \texttt{winning-nonce-found} bit in the status register, and writes the winning nonce to the register file. However, if the corresponding nonce read from the metadata FIFO does not produce a hash that meets the threshold and that nonce corresponds to the maximum value, this implies that the search is over. In that case, the output control \gls{fsm} asserts the \texttt{nonce-not-found} bit in the status register.

\subsection{Chained Hashing Algorithm in Programmable Logic}
\begin{figure*}[t]
  \centering
  \resizebox{0.9\textwidth}{!}{
    \begin{tikzpicture}[%
      font=\scriptsize,%
      inner sep=0pt,%
      minimum width=1.0em,%
      square/.style={regular polygon,regular polygon sides=4},%
      ]
      \tikzset{multiple/.style = {%
          double copy shadow={shadow xshift=.3em,shadow yshift=-.25em,draw=black!60},%
          fill=white,draw=black,thick,minimum height = 2.2em,minimum width=3.1em},%
        ordinary/.style = {rectangle,draw,thick,minimum height = 1cm,minimum width=2cm}}
      
      \def\nsep{0.2}
      \def\fifowidth{1.2em}
      \def\schedwidth{1.2em}

      \node[draw=black,minimum height=3.0em, minimum width=\fifowidth, anchor=north west] (fifo) at (0,0) {\rotatebox{90}{FIFO}};
      \draw[<->] ([yshift=0.6em]fifo.west) -- ++(left:4*\nsep);
      \draw[<-] ([yshift=-0.6em]fifo.west) -- node[above, inner ysep=2pt] {\tiny \texttt{block}} ++(left:4*\nsep);

      \foreach \x in {BLAKE, Keccak, CubeHash, Lyra2, Skein, CubeHash, BMW}
      {
        \node[draw=black,minimum height=3.0em, minimum width=\schedwidth, anchor=north west] (sched) at ($(fifo.north east)+(\nsep,0)$) {\rotatebox{90}{Sched.}};
        \draw[<->] ([yshift=0.6em]sched.west) -- ++(left:\nsep);
        \draw[<-] ([yshift=-0.6em]sched.west) -- ++(left:\nsep);

        \node[multiple, anchor=north] (hash) at ($(sched.south)+(0,-0.25)$) {\x};
        \draw[->] ([xshift=0.3em]hash.north) -- ++(up:.25);
        \draw[<-] ([xshift=-0.3em]hash.north) -- ++(up:.25);

        \node[draw=black,minimum height=3.0em, minimum width=\fifowidth, anchor=north west] (fifo) at ($(sched.north east)+(\nsep,0)$) {\rotatebox{90}{FIFO}};
        \draw[<->] ([yshift=0.6em]fifo.west) -- ++(left:\nsep);
        \draw[<-] ([yshift=-0.6em]fifo.west) -- ++(left:\nsep);
      }
      \node[anchor=east] (rctrl) at ($([yshift=0.6em]fifo.west)-(15*\fifowidth,0)-(13*\nsep,0)$) {};
      \node[anchor=east] (lctrl) at ($(rctrl.center)-(3*\nsep,0)$) {};
      \node (pos) at ($([yshift=0.6em]fifo.east)+(0,2.5*\nsep)$) {};
      \node (pos) at (pos -| lctrl.east) {};
      \draw[<->] ([yshift=0.6em]fifo.east) -- ++(right:2*\nsep) -- ++(up:2.5*\nsep) -- (pos.center);
      \draw[->] ([yshift=-0.6em]fifo.east) -- node[above, inner ysep=2pt] {\tiny \texttt{hash}} ++(right:4*\nsep);

      \draw[densely dashed,color=black!60] ($(lctrl)!0.5!(pos)+(3.5*\nsep,0)$) ellipse (0.07cm and 0.32cm);
      \node[anchor=east] at ($(lctrl)!0.5!(pos)+(2.9*\nsep,0)$) {\tiny \texttt{ctrl}};
      
      \node (pos) at ([yshift=2*-0.25em]hash.south) {};
    \end{tikzpicture}
  }%
  \caption{Architecture of the hardware implementation of the Lyra2REv2 chained hashing algorithm.}
  \label{fig:Lyra2REv2-chain}
\end{figure*}
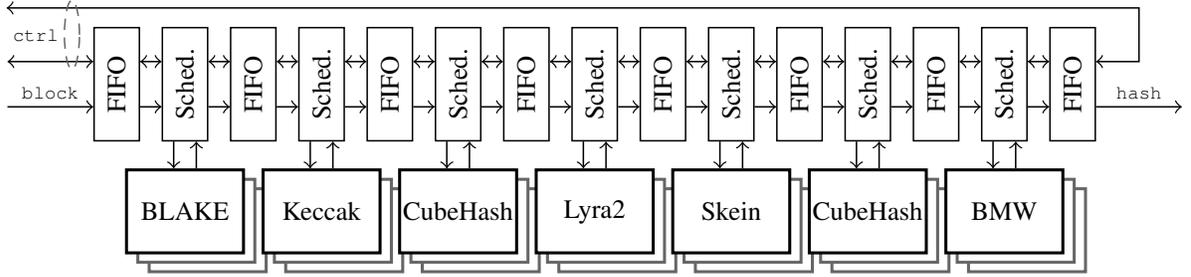

Fig.~\ref{fig:Lyra2REv2-chain} illustrates the hardware implementation of the Lyra2REv2 chained hashing algorithm, where each hash function has its dedicated scheduler, and is bounded by FIFOs. The number of instances of each hash function varies, as it is chosen depending on their respective maximum clock frequency and throughput in hashes per second with the goal to balance the processing pipeline. More details are provided in Section~\ref{sec:results}, but the number of instances per hashing algorithm is selected in order to maximize the overall mining algorithm throughput. This section provides details about the \gls{pl} implementation of the Lyra2REv2 hashing chain.

Reference implementations for the SHA-3 candidates that are optimized for various performance metrics are publicly available. In particular, a research team at the \gls{gmu} described a methodology to compare the hardware performance of fourteen round-two candidates, including all of those utilized in Lyra2REv2 \cite{gmu}, and they also provide the source code for their implementations~\cite{gmusource}. We used the \gls{gmu} throughput-per-area-optimized designs as starting points for some of the implementations of these hashing cores used in this work.

The Lyra2REv2 chain passes only $256$-bit inputs between the algorithms in the chain, while all of the SHA-3 candidates were required to support arbitrary input lengths. Generally, this results in some functionality that does not appear and allows for heavy optimizations. Also, the implementations from \gls{gmu} include interfaces to communicate with software, accounting for such things as endianness and serialization at the output, which are not required for the custom mining chain. As such, we only re-used some main computational blocks of the \gls{gmu} implementations and always customized the control path. This greatly simplifies the control flow for these algorithms and could often also introduce optimizations for the computational datapath. More details are provided for individual hashing cores in the following. 

\subsubsection{FIFOs}
The hashing cores have different nominal frequencies and throughputs. Firstly, FIFOs are used to normalize data transfers between hashing cores with different throughput, by properly asserting the forward- and back-pressure signals. Secondly, since the hashing cores also have various operating frequencies, asynchronous FIFOs are used to safely transfer data from one clock domain to another. The forward and back-pressure signals are individually set to match the internal pipelined architecture of each hashing core.

\subsubsection{Schedulers}
While the FIFOs are necessary to interface hashing algorithms operating at different frequencies, data schedulers---one per hashing step in the chain---are needed to balance throughput between cores with varying execution times. For example, an upstream hashing core producing an output hash every 192 \glspl{cc} will inherently starve a downstream core that can accept new data every 68 \glspl{cc}. To address this limitation, in this example the upstream core would be replicated 3 times and the read/write operation of each core would be scheduled to produce a hash every 64 \glspl{cc}.

The scheduler consists of a state machine that monitors the upstream and downstream FIFO back-pressure signals and that tracks each hashing core computation. Schedulers have knowledge of the execution time and pipeline depth of the hashing cores they are associated to. Given this information, the scheduler will assert the ready signal of the next available core, in a round-robin fashion, when the upstream FIFO has enough data to sustain the hashing core internal pipeline and the downstream FIFO has enough space to receive new data. Subsequently, when a core finishes its computations, the resulting hash is written to the downstream FIFO.

\subsubsection{BLAKE}

Like the round function in the sponge of Lyra2 which is based on BLAKE2b, the round function of BLAKE is given by an arrangement of G-functions. The G-functions themselves differ from the one of Algorithm~\ref{algo:blake2b:g}, with different constants for the rotations and with the insertion of additional adders. We adapted the BLAKE2b round implementation for Lyra2 to implement the BLAKE algorithm.

Consider Fig.~\ref{fig:blake}, which shows the hardware architecture of BLAKE's G-function that updates 4 out of 16 state words, with all signals being $m$ bits wide. In the Lyra2REv2 and Lyra2REv3 algorithms, the BLAKE hash function is for $m=32$\,bits, and uses the constants $R_1 = 16$, $R_2 = 12$, $R_3 = 8$, and $R_4 = 7$. The inputs CM$_{2i}$ and CM$_{2i+1}$ take the value of a round-dependent permutation of a message block $M_n$ and constant $C_o$. Notably, these inputs are excluded when the G-function is implemented together with the sponge, because an interface to inject message blocks into the state is already present in the functions $H.\texttt{absorb}$ and $H.\texttt{duplex}$.

BLAKE hashes a 512-bit message in 14 rounds. In our architecture, which is optimized for high throughput per area, the rounds are fully unrolled and form 14 pipeline stages. The round-dependent permutation can then be designed using only routing resources, rather than requiring a complex block that must be able to output each of the 14 permutations based on a round counter. Furthermore, analogous to Lyra2, the sequential adders within the round are divided into pipeline stages to allow for a higher operating clock frequency of the core. Since the 32-bit adders of BLAKE feature shorter carry chains than the 64-bit adders of Lyra2, only four pipeline stages are implemented within a BLAKE round. In total, the BLAKE architecture forms a 56-stage pipeline that concurrently processes 56 different message blocks. Contrary to the other cores in the Lyra2REv2 chain that pass 256-bit values, the BLAKE core, at the head of the chain, takes 640-bit block headers as input. Each block header is therefore split into two message blocks, and the BLAKE implementation can then, on average, output one hash every 2 \glspl{cc}.

\subsubsection{Keccak} Keccak, which introduced the concept of a sponge, is very efficiently implementable in hardware, which is one of the main reasons it won the SHA-3 competition. While Lyra2 uses a sponge with the BLAKE2b round function, Keccak defines its own family of round functions called \textit{Keccak-f[$w$]}, with $w$ being one of seven values for the sponge permutation width. Lyra2REv2 uses an instance of \textit{Keccak-f[1600]}, with $b+c = 1088 + 512$, the permutation applied in 24 rounds and $l=256$ bits of output hash length. We use a custom sponge implementation with its corresponding control logic, along with the \textit{Keccak-f} round function from \gls{gmu}. Executing one round per clock cycle, the Keccak implementation can then output one hash every 24 \glspl{cc}.

\subsubsection{CubeHash}
Each CubeHash round is simple, but it is applied many times. CubeHash in Lyra2REv2 does $16$ initialization rounds and a total of $176$ finalization rounds. Each round takes a single \gls{cc} so that a total of $192$ \glspl{cc} are required to compute one hash. The difference between initialization and finalization rounds amounts to flipping a single bit of the state and is trivial to implement in hardware. We re-use the CubeHash round function from \gls{gmu}, and implement round-serial control logic to output one hash every 192 \glspl{cc}.

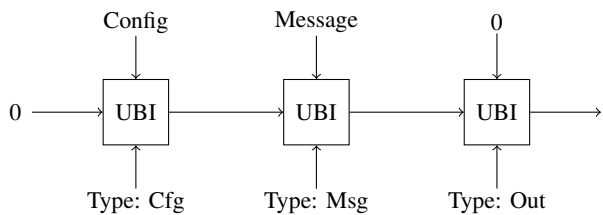
\begin{figure}[t]
  \centering
  \begin{tikzpicture}[font=\small,inner sep=1pt, minimum width=1.2em,square/.style={regular polygon,regular polygon sides=4}]
    \def\nsep{1.6}
    \def\hsep{1.2}

    \node (in) at (0,0) {0};
    \node[draw=black,square] (ubi1) at ($(in)+(\nsep,0)$) {UBI};
    \node[draw=black,square] (ubi2) at ($(in)+(2.5*\nsep,0)$) {UBI};
    \node[draw=black,square] (ubi3) at ($(in)+(4*\nsep,0)$) {UBI};
    \node (out) at ($(in)+(5*\nsep,0)$) {};

    \node (cfg) at ($(in)+(\nsep,\hsep)$) {Config};
    \node (type1) at ($(in)+(\nsep,-\hsep)$) {Type: Cfg};
    \node (msg) at ($(in)+(2.5*\nsep,\hsep)$) {Message};
    \node (type2) at ($(in)+(2.5*\nsep,-\hsep)$) {Type: Msg};
    \node (zero) at ($(in)+(4*\nsep,\hsep)$) {0};
    \node (type3) at ($(in)+(4*\nsep,-\hsep)$) {Type: Out};

    \draw[->] (in) edge (ubi1) (ubi1) edge (ubi2) (ubi2) edge (ubi3) (ubi3) edge (out);
    \draw[->] (cfg) edge (ubi1) (type1) edge (ubi1);
    \draw[->] (msg) edge (ubi2) (type2) edge (ubi2);
    \draw[->] (zero) edge (ubi3) (type3) edge (ubi3);
    
  \end{tikzpicture}
  \caption{Hardware architecture of Skein as a hash function, as depicted in \cite{Ferguson2010}.}%
  \label{fig:skein}%
\end{figure}

\subsubsection{Skein}
Skein is based on the Threefish tweakable block cipher~\cite{Ferguson2010}, and uses the \gls{ubi} chaining mode for hashing, as illustrated in Fig.~\ref{fig:skein}. For the Lyra2REv2 algorithm, all inputs of the first \gls{ubi} block are constant, hence it can be pre-computed as an initialization value. In normal operation of Skein, for an arbitrary length input message, there is an iterative implementation of the \gls{ubi} block, where the last round is slightly different as it inputs the constant zero instead of a message. However, for Skein as used in Lyra2REv2, there is an equal number of hashing rounds (taking message inputs) and finalization rounds (taking zero-inputs). It is useful to unroll and pipeline the remaining two \gls{ubi} blocks. With one of its input as a constant, a significant portion of the logic in the \gls{ubi} block that corresponds to the finalization round can be removed. Furthermore, when using two distinct \gls{ubi} blocks, the key schedule of the first block is independent of the message input, and it can be pre-computed and stored in a read-only memory. Within each \gls{ubi} block, Skein transforms the input using 72 Threefish rounds. Every 8 rounds have a similar structure, and they are implemented as a single pipelined block, which is applied 9 consecutive times. Finally, the above Skein implementation can output, on average, one hash every 9 \glspl{cc}.

\subsubsection{\Glsentrylong{bmw}}
The implementation of \gls{bmw} is derived from that of \gls{gmu}, where the control logic has been completely replaced. \gls{bmw} in Lyra2REv2 takes a 256-bit input and only has a single hash round followed by a finalization round for this input length. Because \gls{bmw} implements fewer rounds, the round function itself is inherently more complex. For example, one of the functions in the round, $f_1$, implements 16 sequential 32-bit adders that each take 17 addition operands. To improve on the achievable clock frequency of the \gls{gmu} design, the round is implemented as 18 pipelined stages. Each message passes once through the pipeline for the hash round and once for the finalization round, such that, on average, the \gls{bmw} implementation outputs one hash at every 2 \gls{cc}.

\subsection{\glsentryshort{mpsoc} Implementation of Lyra2REv3}\label{sec:impl:Lyra2REv3}
A potential \gls{mpsoc} implementation of a Lyra2REv3 miner would be very similar to that of the Lyra2REv2 miner described previously. The required modifications consist of the removal of the Keccak-256 and Skein-256 blocks, the replacement of the simplified Lyra2 block with the new Lyra2MOD block described in Section~\ref{sec:impl:Lyra2MOD}, and the re-arrangement of the hashing chain as shown in Fig.~\ref{fig:Lyra2REv3}.
On the \gls{ps} side, the verification software would need to be modified to use Lyra2REv3, which is also supported by cpuminer-multi.

\begin{table*}[t]
  \centering
  \caption{Throughput metrics for the individual hashing cores for the Xilinx Zynq UltraScale+ \glsentryshort{mpsoc} 9EG.}
  {\begin{tabularx}{.7\textwidth}{l *{6}{|Y} }
    \toprule\hline
    \textbf{Metrics} & BLAKE & Keccak & CubeHash & Lyra2 & Skein & BMW \\
    \hline\hline
    Frequency (MHz)      & 100 & 375 & 250 & 225 & 375 & 100 \\
    Exec. time (\glsentryshortpl{cc}/Hash) & 2 & 24 & 192 & 68 & 9 & 2 \\
    T/P (MHash/s)        & 50.00 & 15.63 & 1.30 & 3.31 & 41.67 & 50.00 \\
    \hline
    \textbf{Combined}\\
    \hline
    \# Cores/Step        & 1 & 2 & 24 & 10 & 1 & 1 \\
    T/P (MHash/s)        & 50.00 & 31.25 & 31.25 & 33.01 & 41.67 & 50.00 \\
    \hline
    \bottomrule
  \end{tabularx}}
  \label{tab:impl:results:metrics}
\end{table*}

\section{Implementation Results}\label{sec:results}
In this section, we provide implementation results for a full standalone Lyra2REv2 miner, notably using the simplified Lyra2 core described in this work.\footnote{We note that the VHDL code and relevant scripts for the simplified Lyra2 core are publicly available at~\url{https://github.com/Michielvb/lyra2-hw}.} To the best of our knowledge, there are no other FPGA-based implementations of simplified Lyra2 cores or for Lyra2REv2 miners in the open literature. For this reason, we can unfortunately not provide detailed comparative FPGA/\gls{mpsoc} implementation results, but we provide a comparison with a GPU and a commercially available FPGA-based Lyra2REv2 miner.

\begin{table*}[t]
  \centering
  \caption{Post-fitting area results of the standalone Lyra2REv2 miner for the Xilinx Zynq UltraScale+ \glsentryshort{mpsoc} 9EG. The average individual results for each hashing core are provided and the total for all combined instances of a core is given in parentheses.}
  \resizebox{\textwidth}{!}{\begin{tabular}{ l *{8}{|c} }
    \toprule\hline
    \textbf{Resources} & BLAKE & Keccak & CubeHash & Lyra2 & Skein & BMW & Others &  \textbf{Total}\\
    \hline\hline
    Area (CLBs)    & \phantom{0}4\,417 (\phantom{0}4\,417) & \phantom{00}\,436 (\phantom{00}\,871) & \phantom{00}\,254 (12\,176) & \phantom{0}1\,206 (12\,062) & \phantom{0}2\,073 (\phantom{0}2\,073) & \phantom{0}2\,064 (\phantom{0}2\,064) & \phantom{0}\,631 & \phantom{0}28\,779 (84\%) \\
    ~~~LUTs        & 25\,229 (25\,229) & \phantom{0}2\,924 (\phantom{0}5\,848) & \phantom{0}1\,762 (84\,553) & \phantom{0}6\,138 (61\,375) & 12\,973 (12\,973) & 12\,153 (12\,153) & 3\,551 & 205\,682 (75\%) \\
    ~~~Registers   & 37\,213 (37\,213) & \phantom{0}2\,013 (\phantom{0}4\,025) & \phantom{0}1\,319 (63\,302) & \phantom{0}8\,321 (83\,211) & 13\,579 (13\,579) & 12\,070 (12\,070) & 3\,667 & 217\,067 (40\%) \\
    ~~~RAM (kbits) & \phantom{00}\,\phantom{00}0 (\phantom{00}\,\phantom{00}0) & \phantom{00}\,\phantom{00}0 (\phantom{00}\,\phantom{00}0) & \phantom{00}\,\phantom{00}0 (\phantom{00}\,\phantom{00}0) & \phantom{0}1\,548 (15\,480) & \phantom{00}\,\phantom{00}0 (\phantom{00}\,\phantom{00}0) & \phantom{00}\,\phantom{00}0 (\phantom{00}\,\phantom{00}0) & 2\,502 & \phantom{0}17\,982 (55\%) \\
    \hline
    \bottomrule
  \end{tabular}}
  \label{tab:impl:results}
\end{table*}

\subsection{Lyra2REv2 Miner}
The Lyra2REv2 miner was implemented on a Xilinx ZCU102 Evaluation Kit, which is based on the Xilinx Zynq UltraScale+ 9EG (ZU9EG) \gls{mpsoc}. The \gls{pl} of the ZU9EG \gls{mpsoc} contains a total of 34\,260 \glspl{clb} with 274\,080 \glspl{lut}, 548\,160 registers, and 32.1\,Mbits of \gls{bram}. The \gls{ps} of the ZU9EG \gls{mpsoc} contains four ARM Cortex-A53 cores clocked at 1.2\,GHz. The functionality of the Lyra2REv2 chain was verified against test vectors generated using cpuminer-multi.

The power-consumption estimation was obtained using Xilinx's Vivado Power Estimator tool, where the timing constraints are those required for the operating frequencies of Table\,\ref{tab:impl:results}, the switching activity is obtained by way of simulation \cite{xilinxsaif} with the miner processing input vectors generated using cpuminer-multi~\cite{cpuminer}, and the post-fitted design provided to the tool meets all timing constraints.

Table~\ref{tab:impl:results:metrics} shows the throughput metrics for the individual hashing cores. Due to the different hashing core architectures, we use a total of 5 clock domains, namely, 100\,MHz for the BLAKE and \gls{bmw} cores, 375\,MHz for Skein and Keccak, 250\,MHz for CubeHash, and 225\,MHz and 450\,MHz for Lyra2 and its multi-pumped RAM blocks, respectively. Clock-domain crossings are done over the asynchronous FIFOs. From Table~\ref{tab:impl:results:metrics}, it can be observed that both the execution time and the resulting individual throughput vary significantly among the hashing cores, thus making it challenging to perfectly balance the Lyra2REv2 chain. The bottom half of Table~\ref{tab:impl:results:metrics} provides the number of cores per hashing step that are used in the Lyra2REv2 chain, which result in a relatively balanced pipeline that is limited by the 31.25\,MHash/s combined throughput of the Keccak and CubeHash cores. It should be noted that there is a total of 48 instances of the CubeHash core as there are two CubeHash steps in the chain (cf. Fig.~\ref{fig:Lyra2REv2-chain}).

Table~\ref{tab:impl:results} shows the post-fitting area results of the proposed Lyra2REv2 miner. Specifically, the table shows the average individual area results for each hashing core and the total amount for all combined instances of a core in parenthesis. The ``Others'' column shows the resource utilization of all blocks except the Lyra2REv2 chain on the PL side of Fig.~\ref{fig:Lyra2REv2-arch}. Finally, the ``Total'' column is the total resource utilization for the complete miner. The total \gls{clb} count is less than the sum of the individual \glspl{clb} because some \glspl{clb} are shared across components. We observe that the 48 CubeHash instances require the most \gls{pl}~\gls{clb} and \gls{lut}~resources, followed closely by the 10 Lyra2 instances. Especially Keccak, on the other hand, is much more hardware efficient.

Table~\ref{tab:impl:cmp} shows the post-fitting power consumption results of the proposed standalone Lyra2REv2 miner. The Lyra2REv2 miner consumes 24.93\,W, which leads to an energy efficiency of 0.80\,$\mu$J/Hash at a throughput of 31.25\,MHash/s.

\begin{table}[t]
  \centering
  \caption{Comparison with a GPU implementation and a commercially available FPGA miner.}
  \begin{tabular}{l | c | c | c }
    \toprule\hline
    \multirow{2}{*}{\bf Implementation} & NVIDIA & Hash Altcoin & Xilinx Zynq\\
    & Titan Xp & BlackMiner F1+ & Ultrascale+ 9EG\\
    \hline\hline
    T/P (MHash/s) 				& 63.09	&  324 & 31.25 \\
    Power (W)     				&  215  &  543 & 25 \\
    En.-Eff. ($\mu$J/Hash)& 3.41  & 1.68 & 0.80 \\
    \hline
    \bottomrule
  \end{tabular}
  \label{tab:impl:cmp}
\end{table}

\subsection{Comparison with a GPU and a Commercial FPGA Miner}
Table~\ref{tab:impl:cmp} shows a performance comparison of the work described in this paper against a Lyra2REv2 miner running on a (non-overlocked) NVIDIA Titan Xp GPU and on the Hash Altcoin BlackMiner F1+ commercial multi-FPGA miner~\cite{hashaltcoin}, which features 18 parallel Xilinx Kintex 7 (XC7K325T) FPGAs. The power consumption of the BlackMiner F1+ has been measured and found to be 543\,W when mining a Lyra2REv2-based cryptocurrency~\cite{blackminerf1preview}. For the GPU, we use version 390.48 of the NVIDIA drivers for Linux and version 2.3.1 of the ccminer software~\cite{ccminer} compiled from scratch with version 9.1.85 of the CUDA compilation tools. The ccminer \texttt{intensity} option was set to 22 (out of 25), which is the largest supported value before the GPU memory runs out. All remaining parameters of the NVIDIA drivers and of the ccminer tool have their default values. We set ccminer up to mine MonaCoin using Lyra2REv2 on the \texttt{zergpool.com} mining pool.\footnote{Note that all mining rewards obtained during testing were directly sent as Vertcoin to the Tip Jar wallet of the Vertcoin Developers (\texttt{VnfNKCy5Aq7vZq5W9UKgMwfDLT7NrPRWZK}), who are also the developers of Lyra2REv2 and Lyra2REv3.} The power and hash rates reported in Table~\ref{tab:impl:cmp} are average values that are provided directly by the ccminer software.

We observe that the proposed FPGA-based Lyra2REv2 miner is estimated to be 4.3 times more energy efficient than the GPU-based miner. Moreover, the FPGA-based Lyra2REv2 miner is also estimated to be 2.1 times more energy efficient than the BlackMiner F1+. Also note that the BlackMiner F1+ is a multi-FPGA miner and that our FPGA-based Lyra2REv2 miner achieves a 1.74 times higher throughput than the average throughput per FPGA of the BlackMiner F1+. However, due to a lack of details on the implementation of the BlackMiner F1+, it is difficult to assess whether the improved energy efficiency and throughput are due to a better implementation of the various hashing cores or simply due to a difference in the employed FPGAs. It should also be noted that the BlackMiner F1+ and our FPGA-based Lyra2REv2 miner are standalone devices, while the power we report for the GPU-based miner does not include the computer required to host the GPU.

\section{Conclusion}\label{sec:conclusion}
This paper, we presented the first FPGA-based implementation of a standalone miner for Lyra2REv2, which is an ASIC-resistant hashing algorithm employed by several cryptocurrencies. To this end, we also presented the first implementation of the simplified Lyra2 hashing algorithm used by Lyra2REv2 in the open literature. The key to achieve a good throughput and energy efficiency for Lyra2 is to efficiently map the memory matrix to \glsfirst{pl} RAM blocks and to pipeline the BLAKE2b round function. With regard to the whole miner, there are two key ingredients. The first one is to minimize communications between software and hardware by implementing nonce generation and threshold verification in hardware. The second one is to optimize the throughput per area of each core in the chain while at the same time finding a good balance between the links, under the constraint of the total amount of resources available. As a result, the proposed Lyra2REv2 FPGA-based miner has an estimated energy efficiency of 0.80\,$\mu$J/Hash at a throughput of 31.25\,MHash/s, which is 4.3 and 2.1 times better than an NVIDIA Titan Xp GPU and a commercial FPGA-based miner, respectively. At the same time, the proposed FPGA-based miner is easily reconfigurable so that it can be adapted to future versions of Lyra2RE which may be introduced to deter ASIC-based miners. Furthermore, with trivial changes to our software, our infrastructure could be reused to mine other cryptocurrencies by swapping cores in the chain.

\section*{Acknowledgment}
The authors gratefully acknowledge the support of NVIDIA Corporation with the donation of a Titan Xp GPU, and of Xilinx for the donation of a Zynq UltraScale+ MPSoC ZCU102 Evaluation Kit. This work was supported in part by the Research Council KU Leuven (C16/15/058), the Horizon 2020 ERC Advanced Grant (695305 Cathedral) and by an NSERC Discovery Launch Supplement (\#651825).


\begin{thebibliography}{10}
\providecommand{\url}[1]{#1}
\csname url@samestyle\endcsname
\providecommand{\newblock}{\relax}
\providecommand{\bibinfo}[2]{#2}
\providecommand{\BIBentrySTDinterwordspacing}{\spaceskip=0pt\relax}
\providecommand{\BIBentryALTinterwordstretchfactor}{4}
\providecommand{\BIBentryALTinterwordspacing}{\spaceskip=\fontdimen2\font plus
\BIBentryALTinterwordstretchfactor\fontdimen3\font minus
  \fontdimen4\font\relax}
\providecommand{\BIBforeignlanguage}[2]{{%
\expandafter\ifx\csname l@#1\endcsname\relax
\typeout{** WARNING: IEEEtran.bst: No hyphenation pattern has been}%
\typeout{** loaded for the language `#1'. Using the pattern for}%
\typeout{** the default language instead.}%
\else
\language=\csname l@#1\endcsname
\fi
#2}}
\providecommand{\BIBdecl}{\relax}
\BIBdecl

\bibitem{VanBeirendonck_ISCAS_2019}
M.~{Van~Beirendonck}, L.-C. Trudeau, P.~Giard, and A.~Balatsoukas-Stimming, ``A
  {Lyra2} {FPGA} core for {Lyra2REv2}-based cryptocurrencies,'' in \emph{{IEEE}
  Int. Symp. on Circuits and Syst. ({ISCAS})}, May 2019.

\bibitem{bitcoin}
S.~Nakamoto, ``Bitcoin: A peer-to-peer electronic cash system,'' 2008.

\bibitem{Dwork1993}
C.~Dwork and M.~Naor, ``Pricing via processing or combatting junk mail,'' in
  \emph{Advances in Cryptology (CRYPTO)}.\hskip 1em plus 0.5em minus
  0.4em\relax Springer Berlin Heidelberg, 1993, pp. 139--147.

\bibitem{monacoin}
\BIBentryALTinterwordspacing
``Mona{C}oin.'' [Online]. Available: \url{https://monacoin.org}
\BIBentrySTDinterwordspacing

\bibitem{verge}
\BIBentryALTinterwordspacing
``Verge.'' [Online]. Available: \url{https://vergecurrency.com}
\BIBentrySTDinterwordspacing

\bibitem{vertcoin}
\BIBentryALTinterwordspacing
``Vertcoin.'' [Online]. Available: \url{http://vertcoin.org}
\BIBentrySTDinterwordspacing

\bibitem{Aumasson2008}
\BIBentryALTinterwordspacing
J.-P. Aumasson, L.~Henzen, W.~Meier, and R.~C.-W. Phan, ``{SHA}-3 proposal
  {BLAKE}, submission to {NIST},'' 2008. [Online]. Available:
  \url{http://131002.net/blake}
\BIBentrySTDinterwordspacing

\bibitem{Bertoni2011}
\BIBentryALTinterwordspacing
G.~Bertoni, J.~Daemen, M.~Peeters, and G.~van Assche, ``The {K}eccak {SHA}-3
  submission,'' 2011. [Online]. Available:
  \url{http://keccak.noekeon.org/Keccak-submission-3.pdf}
\BIBentrySTDinterwordspacing

\bibitem{Ferguson2010}
\BIBentryALTinterwordspacing
N.~Ferguson, S.~Lucks, B.~Schneier, D.~Whiting, M.~Bellare, T.~Kohno,
  J.~Callas, and J.~Walker, ``The {S}kein hash function family,'' 2010.
  [Online]. Available:
  \url{http://www.skein-hash.info/sites/default/files/skein1.3.pdf}
\BIBentrySTDinterwordspacing

\bibitem{Gligoroski2009}
D.~{Gligoroski}, V.~{Klima}, S.~J. {Knapskog}, M.~{El-Hadedy}, and
  J.~{Amundsen}, ``Cryptographic hash function {B}lue {M}idnight {W}ish,'' in
  \emph{Int. Workshop on Security and Commun. Networks}, May 2009, pp. 1--8.

\bibitem{Bernstein2009}
\BIBentryALTinterwordspacing
D.~J. Bernstein, ``Cube{H}ash specification,'' 2009. [Online]. Available:
  \url{http://cubehash.cr.yp.to/submission2/spec.pdf}
\BIBentrySTDinterwordspacing

\bibitem{Tillich2009}
\BIBentryALTinterwordspacing
S.~Tillich, M.~Feldhofer, W.~Issovits, T.~Kern, H.~Kureck,
  M.~M{\"u}hlberghuber, G.~Neubauer, A.~Reiter, A.~K{\"o}fler, and
  M.~Mayrhofer, ``Compact hardware implementations of the {SHA}-3 candidates
  {ARIRANG}, {BLAKE}, {G}r{\o}stl, and {S}kein,'' Cryptology ePrint Archive,
  Report 2009/349, 2009. [Online]. Available:
  \url{https://eprint.iacr.org/2009/349}
\BIBentrySTDinterwordspacing

\bibitem{Baldwin2010}
B.~{Baldwin}, A.~{Byrne}, L.~{Lu}, M.~{Hamilton}, N.~{Hanley}, M.~{O'Neill},
  and W.~P. {Marnane}, ``{FPGA} implementations of the round two {SHA}-3
  candidates,'' in \emph{Int. Conf. on Field Programmable Logic and
  Applications (FPL)}, Aug. 2010, pp. 400--407.

\bibitem{Gaj2010}
K.~Gaj, E.~Homsirikamol, and M.~Rogawski, ``Fair and comprehensive methodology
  for comparing hardware performance of fourteen round two {SHA}-3 candidates
  using {FPGAs},'' in \emph{Cryptographic Hardware and Embedded Systems
  (CHES)}, S.~Mangard and F.-X. Standaert, Eds.\hskip 1em plus 0.5em minus
  0.4em\relax Berlin, Heidelberg: Springer, 2010, pp. 264--278.

\bibitem{gmu}
\BIBentryALTinterwordspacing
E.~Homsirikamol, M.~Rogawski, and K.~Gaj, ``Comparing hardware performance of
  fourteen round two {SHA-3} candidates using {FPGAs},'' Cryptology ePrint
  Archive, Report 2010/445, Dec. 2010. [Online]. Available:
  \url{https://eprint.iacr.org/2010/445}
\BIBentrySTDinterwordspacing

\bibitem{lyra2}
\BIBentryALTinterwordspacing
M.~A. Simpl{\'\i}cio~Jr, L.~C. Almeida, E.~R. Andrade, P.~C. dos Santos, and
  P.~S. Barreto, ``Lyra2: Password hashing scheme with improved security
  against time-memory trade-offs,'' Cryptology ePrint Archive, Report 2015/136,
  2015. [Online]. Available: \url{https://eprint.iacr.org/2015/136}
\BIBentrySTDinterwordspacing

\bibitem{lyra2TC}
E.~R. Andrade, M.~A. Simplicio, P.~S. L.~M. Barreto, and P.~C.~F. d.~Santos,
  ``Lyra2: Efficient password hashing with high security against time-memory
  trade-offs,'' \emph{{IEEE} Trans. Comput.}, vol.~65, no.~10, pp. 3096--3108,
  Oct 2016.

\bibitem{bitcoinreference}
\BIBentryALTinterwordspacing
``Bitcoin developer reference,'' 2019. [Online]. Available:
  \url{https://bitcoin.org/en/developer-reference}
\BIBentrySTDinterwordspacing

\bibitem{lyra2refguide}
M.~A. Simplicio~Jr, L.~C. Almeida, E.~R. Andrade, P.~C. dos Santos, and P.~S.
  Barreto, ``The {Lyra2} reference guide,'' Tech. Report v2.3.2, 2014.

\bibitem{sponge}
G.~Bertoni, J.~Daemen, M.~Peters, and G.~V. Assche, ``Cryptographic sponge
  functions,'' Tech. Report v0.1, Jan. 2011.

\bibitem{sha3}
{NIST}, ``{SHA-3} standard: Permutation-based hash and extendable output
  functions,'' {FIPS} Publication 202, Aug. 2015.

\bibitem{blake2}
J.-P. Aumasson, S.~Neves, Z.~Wilcox-O'Hearn, and C.~Winnerlein, ``{BLAKE2}:
  simpler, smaller, fast as {MD5},'' in \emph{Int. Conf. on Applied Crypto. and
  Netw. Security (ACNS)}.\hskip 1em plus 0.5em minus 0.4em\relax Springer,
  2013, pp. 119--135.

\bibitem{blake}
J.-P. Aumasson, L.~Henzen, W.~Meier, and R.~C.-W. Phan, ``{SHA-3} proposal
  {BLAKE},'' Tech. Report v1.3, Dec. 2010.

\bibitem{Ji2009}
\BIBentryALTinterwordspacing
L.~Ji and X.~Liangyu, ``Attacks on round-reduced {BLAKE},'' Cryptology ePrint
  Archive, Report 2009/238, 2009. [Online]. Available:
  \url{https://eprint.iacr.org/2009/238}
\BIBentrySTDinterwordspacing

\bibitem{vertcoinfork}
\BIBentryALTinterwordspacing
{Vertcoin Development Team Blog}, ``Vertcoin development update,'' Jan. 2019.
  [Online]. Available:
  \url{https://medium.com/vertcoin-blog/vertcoin-development-update-january-2019-8dc39f6df210}
\BIBentrySTDinterwordspacing

\bibitem{Laforest2010}
C.~E. LaForest and J.~G. Steffan, ``Efficient multi-ported memories for
  {FPGAs},'' in \emph{Ann. ACM/SIGDA Int. Symp. on FPGAs (FPGA)}, 2010, pp.
  41--50.

\bibitem{cpuminer}
\BIBentryALTinterwordspacing
T.~Pruvot, ``{cpuminer-multi},'' GitHub repository, 2017. [Online]. Available:
  \url{https://github.com/tpruvot/cpuminer-multi}
\BIBentrySTDinterwordspacing

\bibitem{yocto}
\BIBentryALTinterwordspacing
``Yocto {P}roject.'' [Online]. Available: \url{https://www.yoctoproject.org}
\BIBentrySTDinterwordspacing

\bibitem{gmusource}
\BIBentryALTinterwordspacing
{George Mason University - Cryptographic Engineering Research Group}, ``Source
  code for the {SHA}-3 round 2 candidates \& {SHA}-2 - {H}ash 2011 release.''
  [Online]. Available:
  \url{https://cryptography.gmu.edu/athena/index.php?id=source_codes}
\BIBentrySTDinterwordspacing

\bibitem{xilinxsaif}
\BIBentryALTinterwordspacing
{Xilinx Inc.}, ``{AR\# 53544}: Vivado power analysis - {H}ow do {I} simulate
  for accurate power analysis ({SAIF})?'' [Online]. Available:
  \url{https://www.xilinx.com/support/answers/53544.html}
\BIBentrySTDinterwordspacing

\bibitem{hashaltcoin}
\BIBentryALTinterwordspacing
``{Hash Altcoin BlackMiner F1+},'' 2019. [Online]. Available:
  \url{https://www.hashaltcoin.com/en/batches/11}
\BIBentrySTDinterwordspacing

\bibitem{blackminerf1preview}
\BIBentryALTinterwordspacing
``{Blackminer F1+ Review--FPGA Miner},'' 2019. [Online]. Available:
  \url{https://1stminingrig.com/blackminer-f1-review-fpga-miner/#Verge_XVG_Lyra2rev2_Mining_Hashrate_Power_Draw}
\BIBentrySTDinterwordspacing

\bibitem{ccminer}
\BIBentryALTinterwordspacing
T.~Pruvot, ``{ccminer},'' GitHub repository, 2019. [Online]. Available:
  \url{https://github.com/tpruvot/ccminer}
\BIBentrySTDinterwordspacing

\end{thebibliography}


\begin{IEEEbiography}
	[{\includegraphics[width=1in,height=1.25in,clip,keepaspectratio]{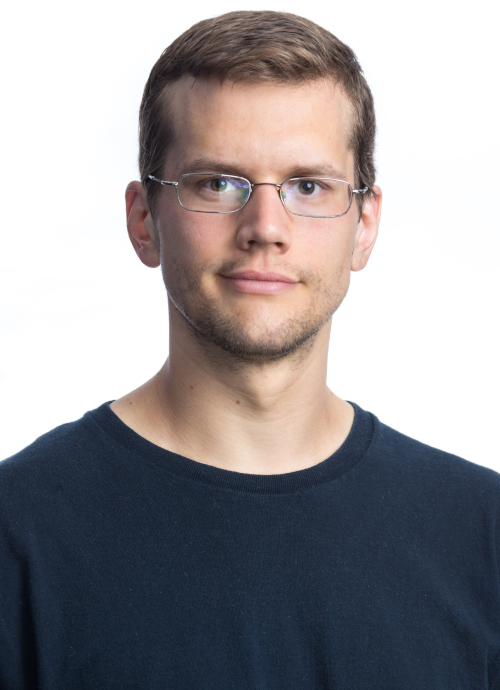}}]{Jean-Fran\c{c}ois T\^{e}tu}
  received the B.Eng. and M.Eng. degrees in Electrical Engineering from \'{E}cole de technologie sup\'erieure (\'{E}TS), Montr\'eal,~Canada, in 2011 and 2014, respectively.
  From 2014 to 2018 he worked in the industry as an embedded software developer, then as an embedded Linux consultant, in Montr\'eal, Canada.
  In late 2018, he then briefly worked as a research professional at the Communications and Microelectronic Integration Laboratory (LaCIME) at \'ETS.
  He is currently a software engineer at a proprietary trading firm in Chicago, USA.
\end{IEEEbiography}
\begin{IEEEbiography}
	[{\includegraphics[width=1in,height=1.25in,clip,keepaspectratio]{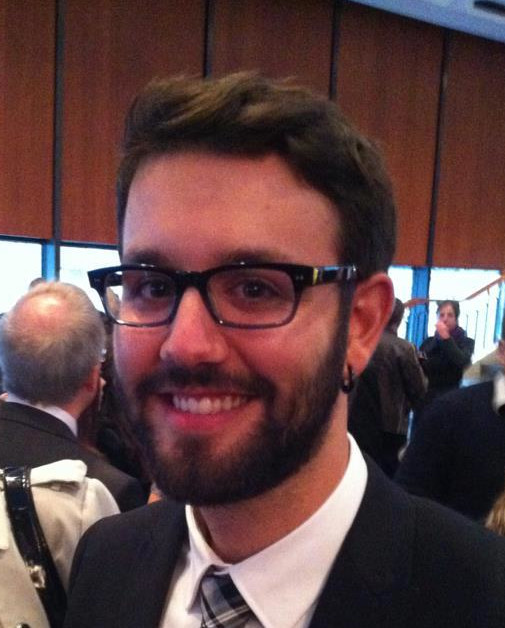}}]{Louis-Charles Trudeau}
  received the B.Eng. and M.Eng. degrees in Electrical Engineering from \'{E}cole de technologie sup\'erieure (\'{E}TS), Montr\'eal,~Canada, in 2012 and 2015, respectively.
  From 2015 to 2018 he worked in the industry as an FPGA designer, specializing in network performance and low-latency financial applications.
  In late 2018, he then briefly worked as a research professional at the Communications and Microelectronic Integration Laboratory (LaCIME) at \'ETS.
  He is currently a hardware engineer at a proprietary trading firm in Chicago, USA.
\end{IEEEbiography}
\begin{IEEEbiography}
	[{\includegraphics[width=1in,height=1.25in,clip,keepaspectratio]{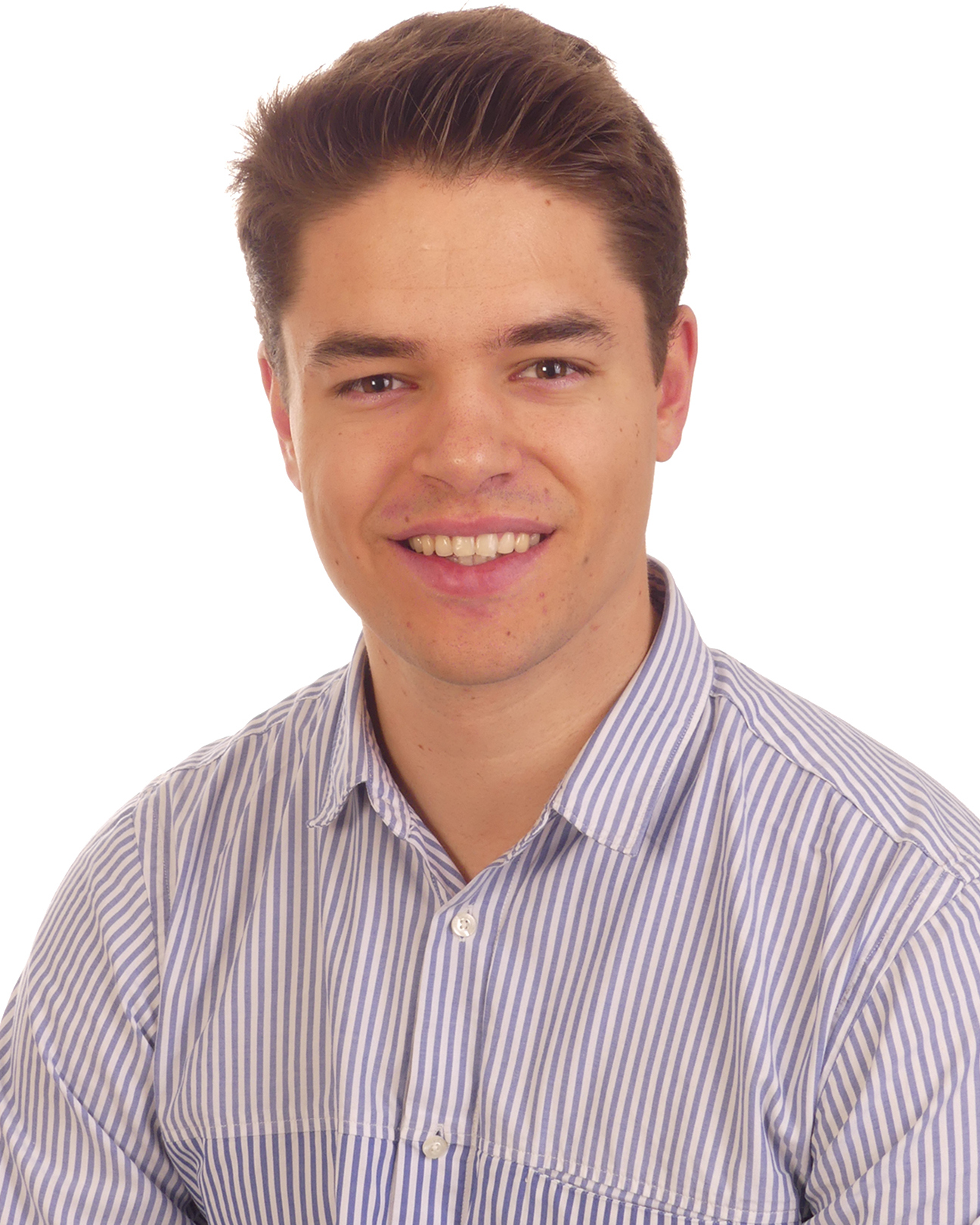}}]{Michiel Van Beirendonck}
  received the B.Sc. and M.Sc. degrees in Electrical Engineering from KU Leuven, Belgium, in 2017 and 2019, respectively. During his M.Sc. studies, he spent one year at EPFL, Switzerland, as part of the SEMP exchange program. Currently, he is pursuing his Ph.D. degree at imec-COSIC, KU Leuven. His research focuses on the implementational challenges of post-quantum cryptography and homomorphic encryption. He is working on efficient architectures that are at the same time able to resist physical attacks.
\end{IEEEbiography}
\begin{IEEEbiography}
	[{\includegraphics[width=1in,height=1.25in,clip,keepaspectratio]{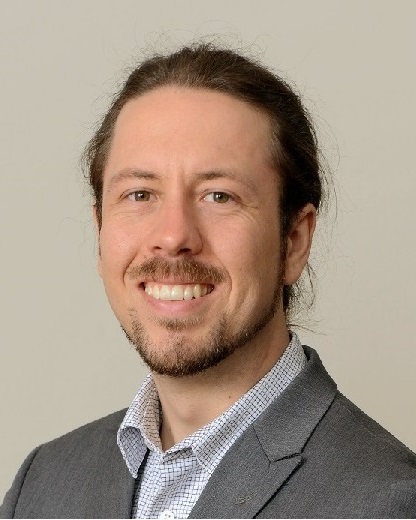}}]{Alexios Balatsoukas-Stimming}
	(S'12--M'17) is currently an Assistant Professor at the Eindhoven University of Technology in the Netherlands. He received the Diploma and MSc degrees in Electronics and Computer Engineering from the Technical University of Crete, Chania, Greece, in 2010 and 2012, respectively, and a PhD in Computer and Communications Sciences from the \'Ecole polytechnique f\'ed\'erale de Lausanne (EPFL), Switzerland, in 2016. He then spent one year at the European Laboratory for Particle Physics (CERN) as a Marie Sk\l{}odowska-Curie postdoctoral fellow and he was a postdoctoral researcher in the Telecommunications Circuits Laboratory at EPFL from 2018 to 2019. His research interests include VLSI circuits for signal processing and communications, error correction coding theory and practice, as well applications of machine learning to signal processing for communications.
\end{IEEEbiography}
\begin{IEEEbiography}[{\includegraphics[width=1in,height=1.25in,clip,keepaspectratio]{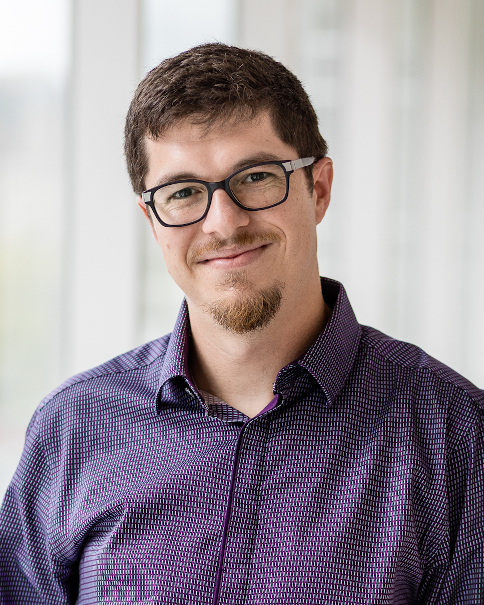}}]{Pascal Giard}
  (S'06--M'10--SM'19) received the B.Eng. and M.Eng. degrees in Electrical Engineering from \'{E}cole de technologie sup\'erieure (\'{E}TS), in 2006 and 2009, respectively, and the Ph.D. degree from McGill University, both in Montr\'eal, Canada.
  From 2009 to 2010, he worked as a research professional in the NSERC -- Ultra Electronics Chair in Wireless Emergency and Tactical Communications at \'ETS. From 2007 to 2016, he was a Lecturer in the Department of Electrical Engineering, \'ETS. He also collaborated as a research professional in the Research Chair in Design Methodology for Highly Integrated and Reliable Hybrid Systems at \'ETS from 2012 to 2016.
  Then, from 2016 to 2018, he was a Postdoctoral Researcher at the Telecommunication Circuits Laboratory, \'Ecole polytechnique f\'ed\'erale de Lausanne (EPFL), Switzerland.
  He is currently an Associate Professor in the Electrical Engineering Department of \'ETS.
His research interests are in the design and implementation of signal processing systems with a focus on modern error-correcting codes, blockchain technology, and connected objects.
He received the Best Experimental-Demonstration Award at the IEEE CASS \& ReSMiQ Innovation Day 2015. He is an Associate Editor for the \textit{Elsevier Microelectronics Journal}.
\end{IEEEbiography}

\end{document}